\documentclass[aps,twocolumn,prd,superscriptaddress,nofootinbib,floatfix]{revtex4-2}

\usepackage{mathtools}
\usepackage{amsfonts}
\usepackage{amssymb}
\usepackage{mathrsfs}
\usepackage{bbm}
\usepackage{slashed}
\usepackage{amsmath}
\usepackage{bm}
\usepackage{bbold}

\usepackage{graphicx}
\usepackage{color}
\usepackage{colortbl}
\usepackage{array}

\usepackage{float}
\usepackage{placeins}
\usepackage{booktabs}
\usepackage[caption=false]{subfig}
\usepackage{makecell}
\usepackage{tabstackengine}

\usepackage{xspace}
\usepackage{hyperref}
\usepackage[nameinlink]{cleveref}
\usepackage{bookmark}
\usepackage{siunitx}

\usepackage{xifthen}
\usepackage{xcolor}
\hypersetup{
	colorlinks,
	linkcolor={red!75!black},
	citecolor={blue!75!black},
	urlcolor={blue!75!black}
}

\usepackage[utf8]{inputenc}
\allowdisplaybreaks[4]

\newcommand{\feyn}[1]{
	\setbox0=\hbox{\ensuremath{#1}}
	\hbox to\wd0{\hbox to0pt{\hbox to\wd0{\hss/\hss}\hss}\box0}}

\def\Eq#1{Eq.~\labelcref{#1}}

\def\Fig#1{Fig.~\labelcref{#1}}

\setkeys{Gin}{width=0.48\textwidth}
\graphicspath{{./figures/}}

\newcommand{\gettitle}{}

\newcommand{\getDalianAffiliation}{\affiliation{School of Physics, Dalian University of Technology, Dalian, 116024, P.R. China}}

\newcommand{\getHeidelbergAffiliation}{\affiliation{Institut f{\"u}r Theoretische Physik, Universit{\"a}t Heidelberg, Philosophenweg 16, 69120 Heidelberg, Germany}}

\hypersetup{
	colorlinks,
	linkcolor={red!75!black},
	citecolor={blue!75!black},
	urlcolor={blue!75!black},
	pdftitle={\gettitle},
	pdfauthor={Zelle},
	pdfkeywords={effective theory} {analytic continuation}
	{correlations functions} {hadronization}
	{functional renormalisation group}
	{real time} {bound states}
	bookmarksopen=true,
	bookmarksopenlevel=2,
	bookmarksnumbered=true
}

\begin{document}

\title{Kaon Distribution Amplitudes from Euclidean Functional QCD}
	
\author{Wen Cui}
\getDalianAffiliation
        
\author{Dao-yu Zhang}
\getDalianAffiliation

\author{Chuang Huang}
\email{huang@thphys.uni-heidelberg.de}
\getHeidelbergAffiliation

\author{Wei-jie Fu}
\getDalianAffiliation

\begin{abstract} 

We study the kaon quasi-distribution amplitude (quasi-DA) and distribution amplitude (DA) within the large-momentum effective theory (LaMET) combined with the first-principles functional QCD. Using quark correlation functions and the kaon Bethe-Salpeter amplitude in the Euclidean space from the 2+1 flavour functional QCD \cite{Fu:2025hcm} as inputs, we obtain the kaon quasi DA in the large longitudinal momentum region with the contour deformation method \cite{Zhang:2025ofc} in the complex plane of momentum. By performing $1/P_z^2$ and $1/P_z^4$ order extrapolations of the kaon quasi-DA for the choices of the maximal longitudinal momentum $P_z^{\max}\in[2,2.5]$ GeV, we obtain a single-peaked and asymmetric kaon DA with the uncertainties arising from the extrapolation interval and ansatz. We find the first and second order moments of the kaon DA, $\langle \xi \rangle_K = 0.020(3)$ and $\langle \xi^2 \rangle_K = 0.253(12)$, respectively.

\end{abstract}
	
\maketitle

\section{Introduction}
\label{sec:int} 

The kaons and pions as Nambu–Goldstone bosons associated with the dynamical chiral symmetry breaking in strong interactions are closely related to the mechanism of mass generation in the Standard Model \cite{Roberts:2021nhw, Ding:2022ows, Raya:2024ejx}. Furthermore, the kaon consists of a strange quark and a light $u$, $d$ quark, which makes the internal structure of the kaon a valuable probe of $SU(3)$ flavour symmetry breaking (FSB) in quantum chromodynamics (QCD). Recently, the BESIII experiment has made significant progress in the study of FSB by measuring the electromagnetic form factors of kaons in the large-momentum-transfer regime \cite{Seth:2013eaa, BESIII:2023zsk}. Experimental results show that the ratio of the electromagnetic form factors of charged kaons to pions exhibits a significant deviation from the leading-order prediction of perturbative QCD \cite{Efremov:1979qk, Farrar:1979aw, Lepage:1979zb, Lepage:1980fj}. In view of the current state of research, the kaon distribution amplitude (DA) plays a key role in the calculation of observables. The nonperturbative kaon DA input has a significant impact on the electromagnetic form factors in the large-momentum-transfer region, see \cite{Gao:2017mmp, LatticeParton:2022zqc, Chen:2024oem, Chai:2025xuz, Ding:2024lfj, Chang:2025lrc}. In recent years, various nonperturbative QCD approaches have achieved significant progress in the calculation of the kaon DA, including the Dyson-Schwinger equation/Bethe-Salpeter equation (DSE/BSE) approach \cite{Shi:2014zpa,Cui:2020tdf,Chang:2025lrc}, holographic models \cite{Swarnkar:2015osa,Chang:2016ouf}, and lattice QCD \cite{LatticeParton:2022zqc, Zhang:2020gaj, RQCD:2019osh}. These approaches generally yield a kaon DA with a slightly narrower, single-peaked structure compared to the pion DA, but there is still no consensus on the extent of the asymmetry in the valence-quark momentum fraction $x$. In this work, we compute the nonperturbative kaon quasi-DA and DA within the large-momentum effective theory (LaMET), where the QCD correlation functions are obtained from first-principles functional QCD calculations.

In the past years, first-principles QCD calculations based on the functional renormalization group (fRG) approach have undergone systematic development, see \cite{Mitter:2014wpa, Braun:2014ata, Rennecke:2015eba, Cyrol:2016tym, Cyrol:2017ewj, Corell:2018yil, Fu:2019hdw, Ihssen:2024miv, Fu:2025hcm, Pawlowski:2025jpg, Fu:2026qnl} and reviews \cite{Dupuis:2020fhh, Fu:2022gou, Rennecke:2025bcw, Fischer:2026uni}. Specifically, the QCD correlation function inputs used in this work are based on a series of papers \cite{Fu:2022uow, Fu:2024ysj, Fu:2025hcm}, where a self-consistent fRG approach to the first-principles 2+1 flavour QCD with controlled systematic uncertainties has been developed, with the only input parameters being the strong coupling and the running quark masses defined at the ultraviolet scale. In this work, we use the Euclidean light- and strange-quark correlation functions and the kaon Bethe–Salpeter amplitude in \cite{Fu:2022uow, Fu:2024ysj, Fu:2025hcm} as the inputs, and combine the LaMET approach to compute the kaon parton distribution amplitude (PDA). In the LaMET approach, the light-cone PDA can be obtained via an extrapolation from the quasi-PDA calculated at a large longitudinal momentum $P_{z}$ in the Euclidean regime, see \cite{Ji:2013dva, Ji:2020ect, Ji:2017rah} for more details. By combining the complex-plane integration method developed in \cite{Zhang:2025ofc}, one is able to perform a direct calculation for the kaon DA based on the first-principles functional QCD in this work. We find for the first two order moments of kaon DA, $\langle\xi\rangle_K = 0.020(3)$ and $\langle\xi^2\rangle_K = 0.253(12)$.

This paper is organized as follows: In \Cref{sec:FunctionalQCD}, we present the truncation setup used in the functional QCD calculations and discuss the quark and kaon-meson correlation functions employed as inputs for the kaon quasi-DA. The introduction of the deformed contour for the momentum integral in the complex plane and its effects will be discussed in \Cref{sec:Deform}, where we also determine a suitable range of large longitudinal momentum for the kaon quasi-DA. In \Cref{sec:PDA}, we present and discuss the final results for the extrapolated light-cone kaon DA and its moments, and compare them with other nonperturbative approaches, showing that our results fall within a reasonable range. We conclude this paper in \Cref{sec:conclusion}.

\section{Correlation functions from functional QCD}
\label{sec:FunctionalQCD}

In \cite{Fu:2025hcm}, the first-principles 2+1 flavour functional QCD framework has been established. By fixing the physical ratios $m_{\pi}/f_{\pi}$ and $m_{K}/f_{\pi}$, this framework requires only the strong coupling and the running quark masses of the light and strange quarks in the ultraviolet as inputs, without introducing any phenomenological parameters or model assumptions. Within the fRG approach, both gluonic and matter degrees of freedom and their dynamics are described self-consistently. In the gluonic sector, the generation of a gluon mass gap leads to its decoupling from the matter sector at low energies, see also \cite{Braun:2007bx, Fischer:2008uz, Fister:2013bh, Mitter:2014wpa, Cyrol:2016tym, Cyrol:2017ewj}. In this section, we focus on the setup and solutions of the quark and meson sectors. The effective action for the quark sector is given by
\begin{align}
    &\quad\Gamma_{\mathrm{quark}}[\bar q, q, A]\nonumber\\[2ex]
    &= \int\limits_p \bar q(-p)\left[\Gamma_{\bar q q}^{(2)}\right](p)\, q(p)+ \int\limits_{p_1,p_2} \left[\Gamma_{A\bar q q}^{(3)}\right]^{a}_{\mu}(p_1,p_2,p_3)\nonumber\\[2ex]
    &\quad\times \bar q (p_2) A^{a}_\mu(p_1) q(p_3)+\sum_\alpha\int\limits_{p_1,p_2,p_3} \left[\Gamma_{4q}^{(4)}\right]^{(\alpha)}_{ijkm}\nonumber\\[2ex]
    &\quad\times \bar q_{i} (p_1) q_{j} (p_2) \bar q_k (p_3) q_m (p_4)\,,
\end{align}
with
\begin{align}
    \int_{p}\equiv\int \frac{d^{4}p}{(2\pi)^{4}}\,.
\end{align}
The label $\alpha$ denotes the different tensor structures for the four-quark vertices. For the two-point function, it reads
\begin{align}
    \Gamma_{\bar q q}^{(2)}(p) = Z_{q}(p)\,\Big[\mathrm{i}\slashed{p}+ M_{q}(p)\Big] \,,
\end{align}
with
\begin{align}
    Z_{q}(p) &= \text{diag}(Z_l(p), Z_l(p), Z_s(p)) \,,\\[2ex]
    M_{q}(p)&=\text{diag}(M_l(p), M_l(p), M_s(p))\,,
\end{align}
where $M_l(p)$ and $M_s(p)$ are the mass functions, and $Z_l(p)$ and $Z_s(p)$ are the wave functions for the light and strange quarks, respectively.

The strength of the quark–gluon interaction plays a crucial role in the quantitative description of chiral symmetry breaking. Therefore, in our calculations, we include not only the classical tensor structure ${\cal T}_{A\bar l l}^{(1)}$ of the quark–gluon vertex, but also the two most important nonclassical channels relevant for the symmetry breaking ${\cal T}_{A\bar l l}^{(4)}$ and ${\cal T}_{A\bar l l}^{(7)}$, enabling a quantitative agreement of physical observables with experiments, where the explicit expressions of tensors above can be found in \cite{Fu:2025hcm}. This is also supported by other nonperturbative QCD studies, see \cite{Fu:2025hcm, Mitter:2014wpa, Williams:2014iea, Williams:2015cvx,  Aguilar:2016lbe, Cyrol:2017ewj, Gao:2021wun, Ihssen:2024miv}. Moreover, in our calculation, all the quark–gluon vertex dressings are momentum dependent, see \cite{Fu:2025hcm} for more details.

%
\begin{figure}[t]
\hspace{-0.4cm}%
\includegraphics[width=0.5\textwidth]{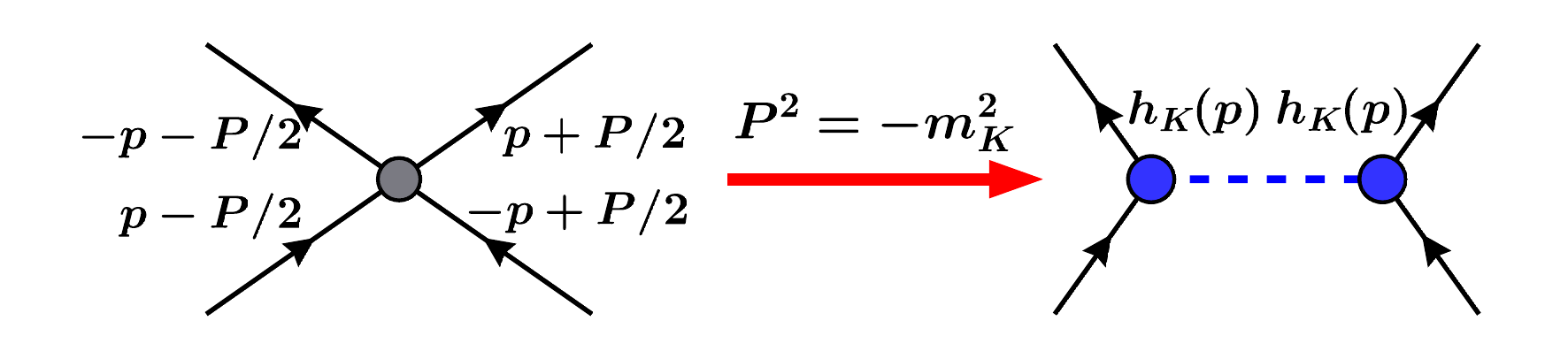}
\caption{Schematic illustration of the decomposition of the four-quark vertex into the two quark-meson vertices connected by a meson propagator, when the $t$-channel momentum is near the pole mass of kaon. Note that the direction of external quark lines denotes that of fermion flows, while the momenta are all counted as incoming towards the vertices.} 
\label{fig:resonanceK}
\end{figure}
%

Within the RG calculations, mesonic degrees of freedom emerge dynamically from four-quark interactions as the RG evolves from the ultraviolet to infrared \cite{Fu:2022uow}. Therefore, for the four-quark interactions we include the four channels $\{\sigma,\,\pi,\,K,\,\kappa\}$, being most relevant for the chiral symmetry breaking. The reliability of this truncation has been validated in \cite{Fu:2022uow, Fu:2024ysj, Fu:2025hcm}. In particular, the four-quark vertex of the kaon channel is given by
\begin{align}
    \left[\Gamma_{4q}^{(4)}\right]_{ijkm}^{(K)}&=(2\pi)^4 \delta^{(4)}\left(\sum_{i=1}^{4}p_i\right)\left(\prod_{i=1}^4 Z_{q}^{\frac{1}{2}}(p_i)\right)  \nonumber\\[2ex]
    &\quad\times \lambda_{K}(p_1,p_2,p_3,p_4){\cal T}^{(K)}_{4q,ijkm}(p_1,p_2,p_3,p_4) \,,
\end{align}
where $\lambda_{K}$ is the dressing of the vertex, and the tensor structure reads,
\begin{align}
    {\cal T}_{4q,ijkm}^{(K)} \bar q_{i} q_{j} \bar q_{k} q_{m} =-\left(\bar{q}\,\gamma_5 T^{(4-7)}q\right)^2\,,
\end{align}
with the Gell-Mann matrices $T^{i}$, the quark field $q=(q_l,q_s)$ and the light quark field $q_l=(q_u,q_d)$. Furthermore, to maximize the extraction of information on the kaon Bethe–Salpeter (BS) amplitude from the four-quark interaction, we employ an $s,t,u$-channel momentum approximation for the dressing, which reads
\begin{align}
 \lambda_K (p_1,p_2,p_3,p_4)&\approx\lambda_K (s,t,u)\,.
\end{align}
Note that the reliability of the $s,t,u$-channel momentum approximation for the four-quark dressings has been investigated and demonstrated in detail in \cite{Fu:2024ysj}. Specifically, the Mandelstam variables are chosen as
\begin{align}
    s=0,\quad t=P^2,\quad u=4p^2\,,
\end{align}
with
\begin{align}
    P=-(p_{1}+p_{2}),\quad p=\frac{p_{2}-p_{1}}{2}=\frac{p_{3}-p_{4}}{2}\,,
\end{align}
where the labels of momenta are shown in \Fig{fig:resonanceK}. We choose
\begin{align}
    P=\sqrt{P^2}\,(1,0,0,0),\quad p=\sqrt{p^2}\,(1,0,0,0)\,,
\end{align}
where the angular dependence is neglected. Such a symmetric momentum choice facilitates the extraction of the BS amplitude, see \Fig{fig:resonanceK}. The BS amplitude is given by the residue of the dressings at the bound-state pole of kaon, which reads
\begin{align}
    h_{K}(p)= \lim_{P^2\to -m_{K}^2} \sqrt{\lambda_{K}(P,p)\cdot(P^2+m_{K}^2)} \,.
\end{align}

%
\begin{figure}[t]
\includegraphics[width=0.45\textwidth]{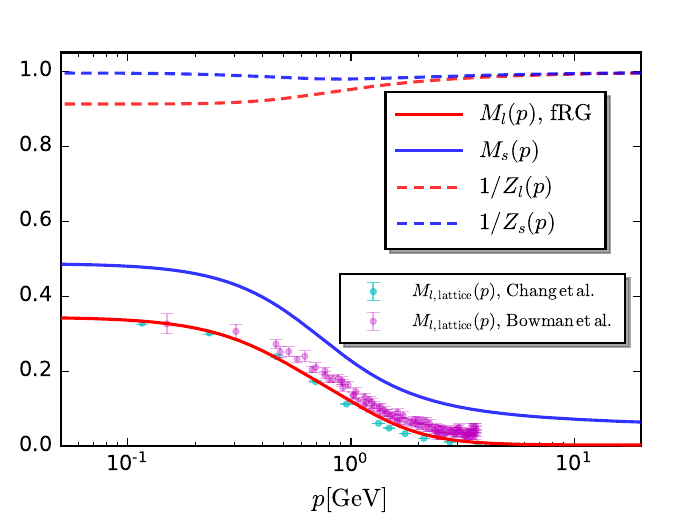}
\caption{Quark mass function $M_{q}(p)$ (solid lines, in unit of GeV) and the quark wave function $Z_{q}(p)$ (dashed lines) for the light (red) and strange (blue) quark, in comparison to the lattice data of light quark mass from \cite{Chang:2021vvx} (cyan points) and \cite{Bowman:2005vx} (purple points).} 
\label{fig:Quark}
\end{figure}
%

%
\begin{figure}[t]
\includegraphics[width=0.45\textwidth]{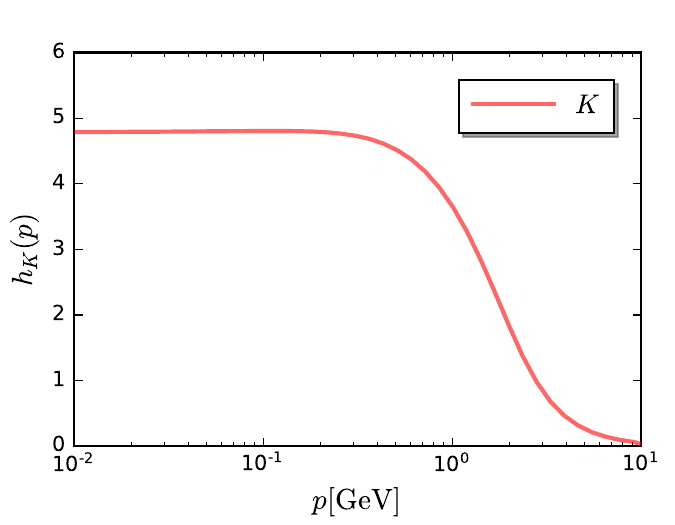}
\caption{Bethe-Salpeter amplitude of the kaon as a function of the quark momentum.} 
\label{fig:BSamplitude}
\end{figure}
%

In \Fig{fig:Quark}, we show the mass and wave functions for the light and strange quarks as functions of the momentum, where the light quark mass function is in agreement with the lattice QCD results \cite{Chang:2021vvx, Bowman:2005vx}. In \Fig{fig:BSamplitude}, the kaon BS amplitude as a function of momentum is shown. It is worth noting that the correlation functions obtained from functional QCD are formulated in Euclidean space. However, the computation of the quasi-DA requires its analytic information in the complex plane. In this work, as in \cite{Zhang:2025ofc}, we use the Taylor expansion in the complex plane of $p_{0}$ to account for their analytic behaviors in the complex plane.  Taking the BS amplitude $h_K (p)$ as an example, one expands it as follows
\begin{align}
    h_{K}(p)& = h_{K}(\bar p)+\sum_{k=1}^{n}\frac{1}{k!}\frac{\partial^k}{\partial p_{0}^k} h_{K}(\bar p)\left(\mathrm{i}\,\Delta p_{0}\right)^k\nonumber\\[2ex]
    &\quad+\mathcal{O}\left(\left(\mathrm{i}\,\Delta p_{0}\right)^{n+1}\right)\,,\label{eq:Taylor-expansion}
\end{align}
with the complex temporal momentum
\begin{align}
    p=(p_{0}+\mathrm{i} \Delta p_{0}, \, p_{3}, \,p_{1}, \, p_{2})\,.
\end{align}
Here, $\bar{p}$ denotes the magnitude of momentum where it is expanded, i.e.,
\begin{align}
    \bar{p}=\sqrt{p_{0}^{2}+p_{1}^{2}+p_{2}^{2}+p_{3}^{2}}\,.
\end{align}
Similarly, we apply the same procedure to the quark mass functions $M_{l}(p)$ and $M_{s}(p)$, and the wave functions $Z_{l}(p)$ and $Z_{s}(p)$ as well. The convergence with respect to the expansion order, as well as its impact on the final DA results, has been discussed in detail in \cite{Zhang:2025ofc, PionPDA:2026}. It was found that the order $n=4$ is sufficient to guarantee the convergence, which is adopted in this work.

\section{Deformed integration contour for kaon quasi-DA}
\label{sec:Deform}

In this section, we discuss the detailed procedure for computing the kaon quasi-DA within the LaMET approach using QCD correlation function inputs shown in \Cref{sec:FunctionalQCD}. The deformed integration contour technique developed in \cite{Zhang:2025ofc} is employed and improved in this work.

We begin with the unamputated Bethe-Salpeter amplitude for the kaon, which is given by
\begin{align}
\chi_K(p;P)=G_l(p_+)\,\Gamma_K(p;P)\,G_s(p_-)\,,\label{eq:unam-BS}
\end{align}
with
\begin{align}
    p_{\pm}=p\pm P/2\,. \label{eq:ppm}
\end{align}
Here, $\Gamma_{K}$ denotes the kaon Bethe–Salpeter amplitude including its tensor structure, 
\begin{align}
    \Gamma_K(p;P)=\mathrm{i}\gamma_5h_K(p;P)\,.
\end{align}
$G_l$ and $G_s$ denote the light- and strange-quark propagators, respectively, which are given by
\begin{align}
    G_l(p_+)&=\frac{-\mathrm{i}\gamma \cdot p_++M_l(p_+)}{Z_{l}(p_{+})\left(p_+^2 +M_l^2(p_+)\right)}, \label{eq:Gl} \\[2ex]
    G_s(p_-)&=\frac{-\mathrm{i}\gamma \cdot p_{-}+M_s(p_-)}{Z_{s}(p_{-})\left(p_{-}^2 +M_s^2(p_{-})\right)}\,.\label{eq:Gs}
\end{align}

In the quasi-light-front framework with the longitudinal momentum, the momentum components of the quark and meson are given by
\begin{align}
    p_\mu &=(p_{0},p_{3},p_{1},p_{2})=(p_{0},p_{3},p_{\perp})\,, \label{eq:momentum-p-lower}\\[2ex]
    P_\mu &=(\mathrm{i}E_{K},\,P_{z},0,0)\,, \label{eq:momentum-p-upper}
\end{align}
with
\begin{align}
    E_{K}=\sqrt{P_{z}^2+m_{K}^{2}}\,.
\end{align}

Subsequently, the kaon quasi-DA is defined as an integral, as follows
\begin{align}
    &\phi_K(x,P_z)\nonumber\\[2ex]
   =&\frac{1}{f_K}\mathrm{Tr}_\mathrm{CD}\bigg[\int\frac{d^4p}{(2\pi)^4}\delta(\tilde{n} \cdot p_+-x\tilde{n} \cdot P)\gamma_5\gamma \cdot \tilde{n}\chi_K(p;P)\bigg]\,,\label{eq:qPDA}
\end{align}
with $\tilde{n}=(0,1,0,0)$, where $f_{K}$ represents the kaon decay constant, and the trace $\mathrm{Tr}_\mathrm{CD}$ works in the color and Dirac spaces. In particular, the integration measure is given by
\begin{align}
    \int\frac{d^4p}{(2\pi)^4}=\frac{1}{(2\pi)^4}\int d p_{0}\int d p_{3} \int d^{2} p_{\perp}\,.
\end{align}
Inserting \Eq{eq:momentum-p-lower} and \Eq{eq:momentum-p-upper} into \Eq{eq:qPDA}, one finds that, in the functional QCD LaMET calculation, the meson on-shell momentum $P^{2}=-m_{K}^{2}$ lies in the complex plane of Minkowski space, while the momentum $p$ in the integrand are defined in Euclidean space. Thus, when we do the integration over $p_{0}$, the poles of the unamputated Bethe–Salpeter amplitude might cross the real axis for some range of $x$, leading to the breakdown of the calculations. Specifically, using the quark propagators in \Eq{eq:Gl} and \Eq{eq:Gs} and assuming constant quark masses, one is able to obtain four different poles of the integrand as follows
\begin{align}
     p_{0,\mathrm{pole}\,1}&=\mathrm{i}\bigg[-\frac{\sqrt{P_z^2+m_K^2}}{2}-\sqrt{p_\perp^2+(xP_z)^2+M_l^2}\bigg]\,,\nonumber\\[1ex]
     p_{0,\mathrm{pole}\,2}&=\mathrm{i}\bigg[-\frac{\sqrt{P_z^2+m_K^2}}{2}+\sqrt{p_\perp^2+(xP_z)^2+M_l^2}\bigg]\,,\nonumber\\[1ex]
     p_{0,\mathrm{pole}\,3}&=\mathrm{i}\bigg[\frac{\sqrt{P_z^2+m_K^2}}{2}-\sqrt{p_\perp^2+\big((x-1)P_z \big)^2+M_s^2}\bigg]\,,\nonumber\\[1ex]
     p_{0,\mathrm{pole}\,4}&=\mathrm{i}\bigg[\frac{\sqrt{P_z^2+m_K^2}}{2}+\sqrt{p_\perp^2+\big((x-1)P_z \big)^2+M_s^2}\bigg]\,,\label{eq:quark-poles}
\end{align}
two of which, i.e., $p_{0,\mathrm{pole}\,2}$ and $p_{0,\mathrm{pole}\,3}$, might cross the real axis and enter into the other half-plane. In \cite{Zhang:2025ofc}, a finite imaginary shift $p_{0}\to p_{0}+\mathrm{i}C$ in the integral of $p_{0}$ is introduced to circumvent this difficulty, that is
\begin{align}
    \int_{-\infty}^{\infty}d p_0 \to \int_{-\infty +\mathrm{i} C}^{\infty +\mathrm{i} C}d p_0\,,\label{eq:integral-shift}
\end{align}
The shift quantity $C$ is chosen such that the shifted integral contour crosses between the first and second poles and the third and fourth poles, see \cite{Zhang:2025ofc} for more details.  To that end, in this work, we choose
\begin{align}
C = 
\begin{cases}
g(0), & x < 0 \\[1ex]
g(x), & 0 \le x \le 1 \\[1ex]
g(1), & x > 1
\end{cases}\,,
\end{align}
with the function
\begin{align}
g(x)=\frac{\mathrm{Im}\,p_{0,\mathrm{pole}\,2}+\mathrm{Im}\,p_{0,\mathrm{pole}\,3}}{2}\,.
\end{align}

%
\begin{figure}[t]
\includegraphics[width=0.45\textwidth]{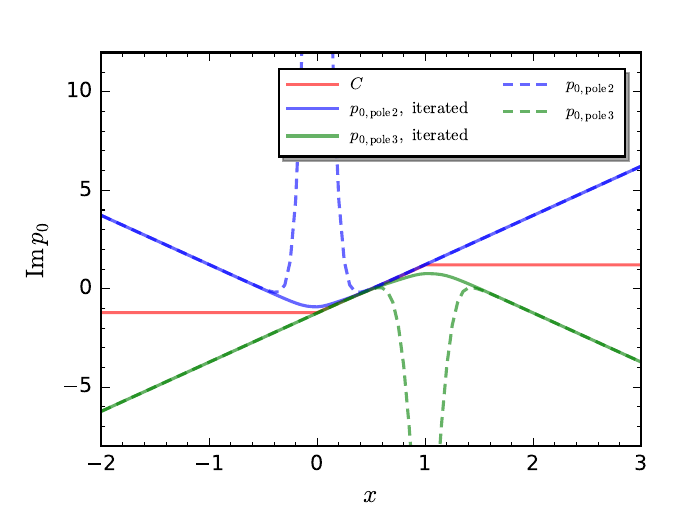}
\caption{Imaginary parts of the second and third poles in Eqs.~\labelcref{eq:quark-poles} as functions of the momentum fraction $x$ with $p_{\perp}=0$, $P_{z}=2.5$ GeV, and $m_K=0.49$ GeV. The red solid line denotes the magnitude of the shift in the imaginary axis in the integral of $p_{0}$ for the quasi-DA. The blue and green solid lines denote the imaginary parts of $p_{0,\mathrm{pole}\,2}$ and $p_{0,\mathrm{pole}\,3}$ as functions of $x$ after the iterative procedure, respectively. The blue and green dashed lines denote the imaginary parts of $p_{0,\mathrm{pole}\,2}$ and $p_{0,\mathrm{pole}\,3}$ as functions of $x$ without the iterative procedure, respectively.} 
\label{fig:pole}
\end{figure}
%

%
\begin{figure}[t]
\includegraphics[width=0.45\textwidth]{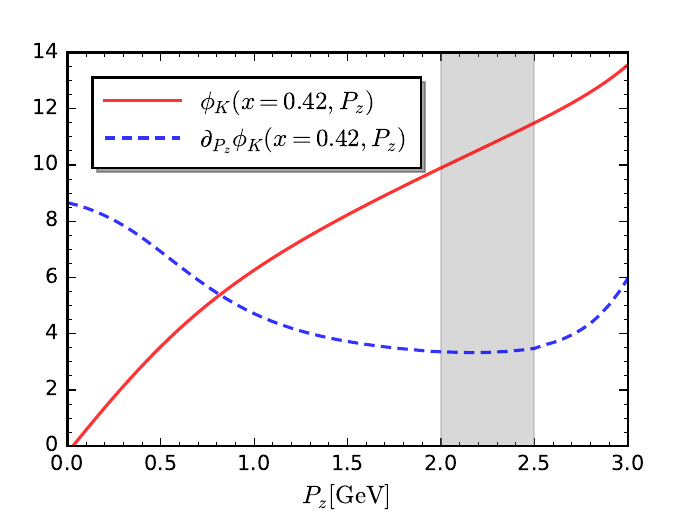}
\caption{Non-normalized quasi-PDA $\phi_{K}(x,P_{z})$ (red solid) and its derivative $\partial \phi_{K}(x,P_{z})/\partial P_{z}$ (blue dashed) at $x=0.42$ as a function of the longitudinal momentum $P_{z}$. The gray band $P_{z}\in [2,2.5]$ GeV indicates the region where the first derivative of the quasi-DA attains its minimum.} 
\label{fig:Pz-choice}
\end{figure}
%

It should be noted that, as shown in Eqs.~\labelcref{eq:quark-poles}, the right-hand side of the equations there are the quark mass functions $M_{l}(p)$ and $M_{s}(p)$, which also depend on the shifted momentum $p_0 + \mathrm{i}C$. Therefore, in practical calculations, the poles and the shift are determined through an iterative procedure until a stable convergence is reached. In \Fig{fig:pole}, we present the results of the integration shift and the pole results with and without iterations. It is observed that, after the iterative procedure, the divergent peaks of the poles in the endpoint region of $x$ are removed. This improves on the stability of the quasi-DA calculation and extends the accessible range of longitudinal momentum.

After the shift, the maximal accessible range of the longitudinal momentum $P_{z}$ is determined by the crossing positions of $p_{0,\mathrm{pole}\,2}$ and $p_{0,\mathrm{pole}\,3}$, which require
\begin{align}
    \mathrm{Im}\,p_{0,\mathrm{pole}\,2}>\mathrm{Im}\,\,p_{0,\mathrm{pole}\,3}\,,
\end{align}
that is
\begin{align}
    M_l^2+M_s^2+2p_\perp^2+2\sqrt{p_\perp^2+M_l^2}\sqrt{p_\perp^2+M_s^2}>m_K^2\,.\label{eq:bound-condition}
\end{align}
If the quark masses are treated as constants, \Eq{eq:bound-condition} is automatically satisfied. However, once momentum-dependent quark mass functions are taken into account, it may no longer be valid. In particular, after performing the $p_{3}$ integration via the delta function in \Eq{eq:qPDA}, which brings a dependence on $P_{z}$ in the quark mass functions, the allowed range of $P_{z}$ is constrained by \Eq{eq:bound-condition}, see also \cite{Zhang:2025ofc} for more details.

In \Fig{fig:Pz-choice}, we show the kaon quasi-DA $\phi_{K}(x, P_{z})$ and its derivative $\partial \phi_{K}(x,P_{z})/\partial P_{z}$ at $x = 0.42$ as functions of $P_z$. We find that the second and third poles are closest to each other at $x = 0.42$, which also corresponds to the peak position of the quasi-DA. In this work, the longitudinal momentum can be extended to values up to about 3 GeV. However, unlike the pion results reported in \cite{Zhang:2025ofc, PionPDA:2026}, no plateau region or saturation behavior is observed within the present accessible range of $P_{z}$, which warrants further investigation in future work. The grey area in \Fig{fig:Pz-choice} shows that the first derivative of the quasi-DA with respect to $P_z$ attains its minimum in the flat region $P_z \in [2, 2.5]$ GeV. In calculations, this interval is used to determine the maximal momentum $P_z^{\mathrm{max}}$ for the longitudinal momentum region employed to extrapolate the light-cone kaon PDA. This constitutes one of the sources of systematic uncertainties in our calculation.

%
\begin{figure}[t]
\includegraphics[width=0.45\textwidth]{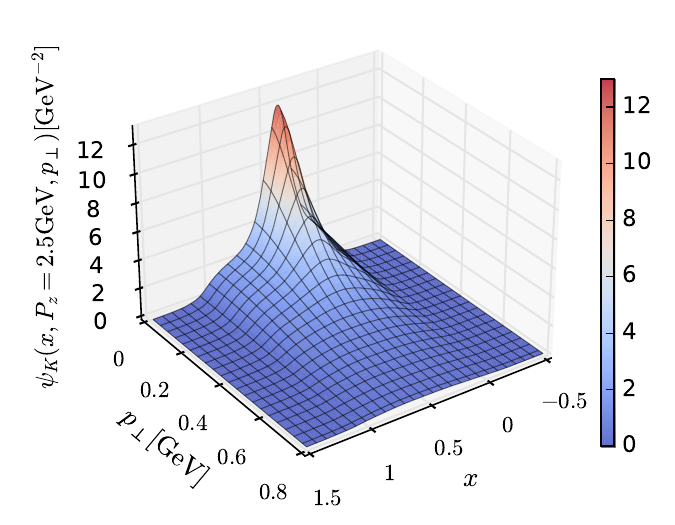}
\caption{3D plot of the kaon quasi-LFWF $\psi_{K}$ as a function of the momentum fraction $x$ and transverse momentum $p_\perp$ at the longitudinal momentum $P_z$ = 2.5 GeV.} 
\label{fig:LFWF}
\end{figure}
%

We conclude this section with the results for the kaon quasi-light-front wave function (quasi-LFWF) presented in \Fig{fig:LFWF}, which is defined as
\begin{align}
    &\quad \psi_K(x,P_z,p_{\perp})\nonumber\\[2ex]
    &\hspace{-0.15cm}=\frac{1}{f_K}\mathrm{Tr}_\mathrm{CD}\bigg[\int\frac{d p_{0} d p_{3}}{(2\pi)^2}\delta(\tilde{n} \cdot p_+-x\tilde{n} \cdot P)\gamma_5\gamma \cdot \tilde{n}\chi_K(p;P)\bigg]\,.\label{eq:qLFWF}
\end{align}
We find that the results of the quasi-LFWF already exhibit an asymmetry in $x$. By further integrating over the transverse momentum $p_{\perp}$, the kaon quasi-DA is obtained.

\section{Kaon distribution amplitude and its moments}
\label{sec:PDA}

%
\begin{figure}[t]
\includegraphics[width=0.45\textwidth]{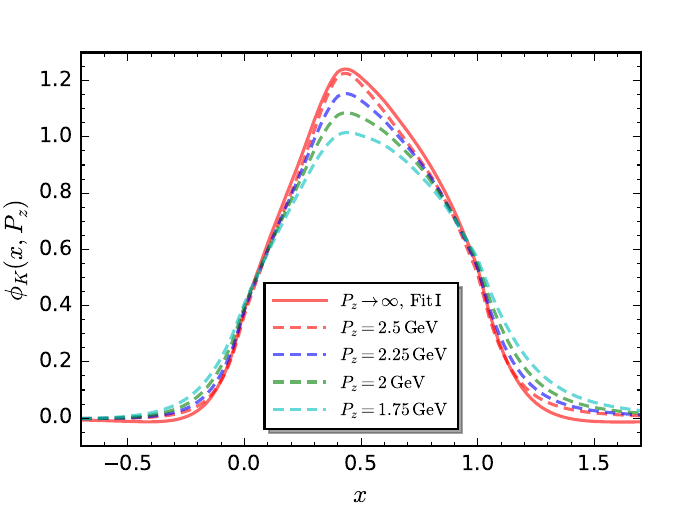}
\caption{Kaon quasi-DAs as functions of the momentum fraction $x$ calculated with several different finite values of $P_z$ from 1.75 GeV to 2.5 GeV. The red solid line denotes the extrapolation result of order $1/P_z^{2}$ with the maximal momentum of the extrapolation regime set to $P_z^{\mathrm{max}}=2.5$ GeV.} 
\label{fig:quasi-pda}
\end{figure}
%

%
\begin{figure}[t]
\includegraphics[width=0.45\textwidth]{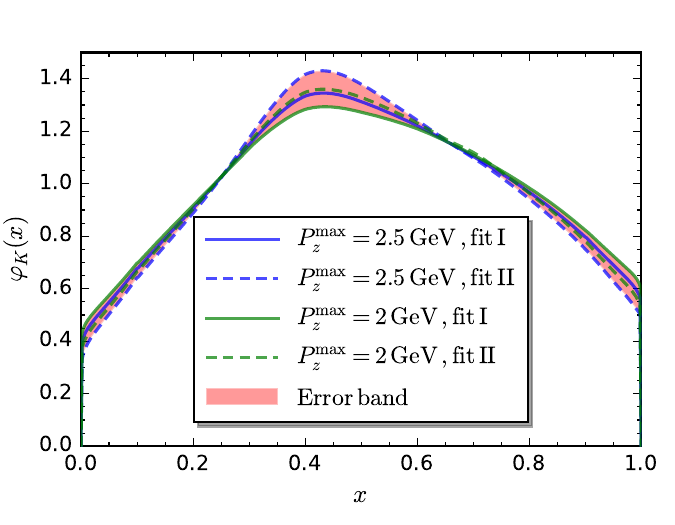}
\caption{Light-front PDA of kaon as a function of the momentum fraction $x$. For the maximal extrapolation range $P_z^{\mathrm{max}}$ set to 2.5 GeV, the blue solid line denotes the result of order $1/P_{z}^{2}$ extrapolation in \Eq{eq:PDA-fit1}, the blue dashed line denotes the result of order $1/P_{z}^{4}$ extrapolation in \Eq{eq:PDA-fit2}. The green lines correspond to the results with the maximal $P_z$ extrapolation range set to 2 GeV. The error band originates from the choice of the $P_z$ extrapolation interval and ansatz, as shown in \Fig{fig:Pz-choice}, \Eq{eq:PDA-fit1} and \Eq{eq:PDA-fit2}. } 
\label{fig:pda}
\end{figure}
%

In this section, we discuss the extrapolation of the kaon quasi-DA, analyze the uncertainties, and present the final light-cone kaon DA. After that we will compare its first two order moments with results from Lattice QCD and DSE/BSE methods.

In \Fig{fig:quasi-pda}, we show the kaon quasi-DA for $P_z$ ranging from 1.75 GeV to 2.5 GeV. From the results, the tilt of the peak structure has already indicated the flavour symmetry breaking in the internal structure of kaon. The quasi-DA becomes increasingly narrower as $P_z$ increases, particularly in the vicinity of the peak region. Different from the pion results within the functional LaMET framework \cite{Zhang:2025ofc,PionPDA:2026}, no saturation behavior of the kaon quasi-DA is observed at large $P_z$ in the present calculation. This necessitates a more careful assessment of the uncertainties in the $P_z$ extrapolation. In this work, we adopt two extrapolation schemes in powers of $1/P_z$, up to second and fourth order, respectively, which are given by
\begin{align}
    \phi_K(x, P_z)=\phi_K^{\mathrm{fit\,I}}(x, P_z\to \infty)+\frac{c_2(x)}{P_z^2}+\mathcal{O}\left(\frac{1}{P_z^4}\right)\,,\label{eq:PDA-fit1}
\end{align}
and
\begin{align}
    &\quad\phi_K(x, P_z)\nonumber\\[2ex]
    &=\phi_K^{\mathrm{fit\,II}}(x, P_z\to \infty)+\frac{c_2(x)}{P_z^2}+\frac{c_4(x)}{P_z^4}+\mathcal{O}\left(\frac{1}{P_z^6}\right)\,.\label{eq:PDA-fit2}
\end{align}
We find that further increase of the order of $P_z$ expansion has a negligible effect on the extrapolated results. The extrapolation uncertainty considered here, together with the uncertainty associated with the choice of the $P_z$ extrapolation range in \Cref{sec:Deform}, constitutes the systematic uncertainty in the light-cone kaon DA. The extrapolated results in \Fig{fig:quasi-pda} are close to the quasi-DA results at $P_z = 2.5$ GeV, which indicates that our choice of the extrapolation interval is reasonable.

Due to the limitations of the LaMET approach in the small-$x$ region, an additional fit of $\phi_K(x, P_z\to \infty)$ in the endpoint regions of $x$ is required to obtain the kaon DA $\varphi_{K}(x)$. It reads
\begin{align}
    \varphi_{K}(x)= c\,x^{a}(1-x)^{b}\,,\label{eq:x-extrap}
\end{align}
in the regime
\begin{align}
0<x<x_{\mathrm{EP}}\,\,\,\,\text{and}\,\,\,1-x_{\mathrm{EP}}<x<1\,.
\end{align}
By considering different fitting intervals, we find that the fitting remains very stable and does not affect the shape or the moments of the kaon DA. This is also supported by the pion studies in \cite{Zhang:2025ofc, PionPDA:2026}. In this work, we take $x_{\mathrm{EP}}=0.05$.

The final result for the kaon DA is shown in \Fig{fig:pda}. The kaon DA shows a single-peaked and asymmetric structure, with its peak located at $x = 0.42$. As the $P_z$ fitting range is increased and higher-order extrapolation functions are employed, the DA becomes relatively narrower. The $1/P_{z}^{2}$ extrapolation with $P_z^{\mathrm{max}}$=2.5 GeV is in close agreement with the $1/P_{z}^{4}$ extrapolation with $P_z^{\mathrm{max}}$=2 GeV, and falls into the central region of the uncertainty band. This indicates that the kaon quasi-DA possesses a reliable extrapolation limit. In \Cref{tab:moments}, we present the first and second moments of the kaon DA and provide a comparison with other approaches \cite{Chang:2025lrc, LatticeParton:2022zqc, RQCD:2019osh}. The moments of DA are defined as
\begin{align}
    \langle\xi^n\rangle_K = \int_{0}^{1} dx\, (2x-1)^n \varphi_K(x)\,.
\end{align}
The first moment quantifies the asymmetry of the DA, which is associated with flavour symmetry breaking. The second moment characterizes the width of the DA, which is directly related to the dynamical chiral symmetry breaking, see also \cite{Chang:2013pq, Cui:2020tdf}. It is found that our result for the second moment is consistent with other approaches, especially lattice QCD LaMET results. The first moment obtained in this work is smaller compared to other approaches but remains within the uncertainty range, which calls for further investigation in future studies.

%
\begin{table}[t]
  \centering
  \begin{tabular}{c||c|c}
    \hline\hline & &    \\[-2ex]   
       & $\langle\xi\rangle_K$ & $\langle\xi^{2}\rangle_K $ \\[1ex]
      \hline & &   \\[-2ex]
      \,\,This work\,\, & \,\,0.020(3)\,\, &  \,\,0.253(12)\,\,   \\[1ex]
      \hline & &   \\[-2ex]
      \,\,DSE/BSE \cite{Chang:2025lrc}\,\, & \,\,0.082(7)\,\, &  \,\,0.239(9)\,\,   \\[1ex]
      \hline & &   \\[-2ex]
      \,\,Lattice LaMET \cite{LatticeParton:2022zqc}\,\, & \,\,0.065(31)\,\, &  \,\,0.258(32)\,\,   \\[1ex]
      \hline & &   \\[-2ex]
      \,\,Lattice OPE (RQCD) \cite{RQCD:2019osh}\,\, & \,\,0.032(12)\,\, &  \,\,0.231(4)\,\,   \\[1ex]
      \hline\hline
  \end{tabular}
  \caption{The first and second order moments of the valence-quark parton distributions of the kaon in this work. The results are compared with the results from DSE/BSE computation \cite{Chang:2025lrc} and different Lattice QCD computations \cite{LatticeParton:2022zqc, RQCD:2019osh}. }
  \label{tab:moments}
\end{table}
%

\section{Conclusions and Outlook}
\label{sec:conclusion}

In this work, based on a first-principles functional QCD framework combined with the LaMET method, we study the kaon quasi-DA at large longitudinal momentum and DA in the light-cone. We employ the quark correlation functions and the kaon Bethe–Salpeter amplitude obtained from the 2+1 flavour QCD computation within the fRG approach as inputs to construct the kaon quasi-DA. By performing integration contour deformation in the complex plane, we achieve a maximal accessible longitudinal momentum up to 3 GeV. For stability, we restrict the maximal momentum of the extrapolation range to $P_z^{\mathrm{max}} \in [2, 2.5]$ GeV.

By performing $1/P_z^2$ and $1/P_z^4$ order extrapolations together with endpoint fittings, we obtain the kaon light-cone DA and quantify the associated systematic uncertainties. The resulting DA is single-peaked and asymmetric, with its maximum located at $x = 0.42$. We further obtain $\langle \xi \rangle_K = 0.020(3)$ and $\langle \xi^2 \rangle_K = 0.253(12)$. The second moment shows good agreement with lattice QCD \cite{LatticeParton:2022zqc, RQCD:2019osh} and DSE/BSE results \cite{Chang:2025lrc}, indicating a reliable description of the DA width, while the first moment is slightly smaller but still consistent within current uncertainties.

In the present calculation, we do not observe a clear saturation behavior of the kaon quasi-DA in the large $P_z$ region. This indicates that, for heavier mesons, further development of direct Minkowski-space computations within functional QCD is required, as well as improvements on the extrapolation procedures within the current LaMET framework. In addition, we plan to extend our studies to other observables, such as parton distribution functions and form factors, in future work.

\section*{Acknowledgements}

We thank Yang-yang Tan for discussions. This work is supported by the National Natural Science Foundation of China under Grant No.\ 12447102, and the Collaborative Research Centre SFB 1225 (ISOQUANT).



\bibliography{ref-lib}%

\begin{thebibliography}{53}%
\makeatletter
\providecommand \@ifxundefined [1]{%
 \@ifx{#1\undefined}
}%
\providecommand \@ifnum [1]{%
 \ifnum #1\expandafter \@firstoftwo
 \else \expandafter \@secondoftwo
 \fi
}%
\providecommand \@ifx [1]{%
 \ifx #1\expandafter \@firstoftwo
 \else \expandafter \@secondoftwo
 \fi
}%
\providecommand \natexlab [1]{#1}%
\providecommand \enquote  [1]{``#1''}%
\providecommand \bibnamefont  [1]{#1}%
\providecommand \bibfnamefont [1]{#1}%
\providecommand \citenamefont [1]{#1}%
\providecommand \href@noop [0]{\@secondoftwo}%
\providecommand \href [0]{\begingroup \@sanitize@url \@href}%
\providecommand \@href[1]{\@@startlink{#1}\@@href}%
\providecommand \@@href[1]{\endgroup#1\@@endlink}%
\providecommand \@sanitize@url [0]{\catcode `\\12\catcode `\$12\catcode
  `\&12\catcode `\#12\catcode `\^12\catcode `\_12\catcode `\%12\relax}%
\providecommand \@@startlink[1]{}%
\providecommand \@@endlink[0]{}%
\providecommand \url  [0]{\begingroup\@sanitize@url \@url }%
\providecommand \@url [1]{\endgroup\@href {#1}{\urlprefix }}%
\providecommand \urlprefix  [0]{URL }%
\providecommand \Eprint [0]{\href }%
\providecommand \doibase [0]{https://doi.org/}%
\providecommand \selectlanguage [0]{\@gobble}%
\providecommand \bibinfo  [0]{\@secondoftwo}%
\providecommand \bibfield  [0]{\@secondoftwo}%
\providecommand \translation [1]{[#1]}%
\providecommand \BibitemOpen [0]{}%
\providecommand \bibitemStop [0]{}%
\providecommand \bibitemNoStop [0]{.\EOS\space}%
\providecommand \EOS [0]{\spacefactor3000\relax}%
\providecommand \BibitemShut  [1]{\csname bibitem#1\endcsname}%
\let\auto@bib@innerbib\@empty
\bibitem [{\citenamefont {Fu}\ \emph {et~al.}(2025)\citenamefont {Fu},
  \citenamefont {Huang}, \citenamefont {Pawlowski}, \citenamefont {Tan},\ and\
  \citenamefont {Zhou}}]{Fu:2025hcm}%
  \BibitemOpen
  \bibfield  {author} {\bibinfo {author} {\bibfnamefont {W.-j.}\ \bibnamefont
  {Fu}}, \bibinfo {author} {\bibfnamefont {C.}~\bibnamefont {Huang}}, \bibinfo
  {author} {\bibfnamefont {J.~M.}\ \bibnamefont {Pawlowski}}, \bibinfo {author}
  {\bibfnamefont {Y.-y.}\ \bibnamefont {Tan}},\ and\ \bibinfo {author}
  {\bibfnamefont {L.-j.}\ \bibnamefont {Zhou}},\ }\bibfield  {title} {\bibinfo
  {title} {{Four-quark scatterings in QCD III}},\ }\href
  {https://doi.org/10.1103/4sh5-w4yc} {\bibfield  {journal} {\bibinfo
  {journal} {Phys. Rev. D}\ }\textbf {\bibinfo {volume} {112}},\ \bibinfo
  {pages} {054047} (\bibinfo {year} {2025})},\ \Eprint
  {https://arxiv.org/abs/2502.14388} {arXiv:2502.14388 [hep-ph]} \BibitemShut
  {NoStop}%
\bibitem [{\citenamefont {Zhang}\ \emph {et~al.}(2025)\citenamefont {Zhang},
  \citenamefont {Huang},\ and\ \citenamefont {Fu}}]{Zhang:2025ofc}%
  \BibitemOpen
  \bibfield  {author} {\bibinfo {author} {\bibfnamefont {D.-y.}\ \bibnamefont
  {Zhang}}, \bibinfo {author} {\bibfnamefont {C.}~\bibnamefont {Huang}},\ and\
  \bibinfo {author} {\bibfnamefont {W.-j.}\ \bibnamefont {Fu}},\ }\bibfield
  {title} {\bibinfo {title} {{Quasi parton distributions of pions at large
  longitudinal momentum}},\ }\href@noop {} {\  (\bibinfo {year} {2025})},\
  \Eprint {https://arxiv.org/abs/2502.15384} {arXiv:2502.15384 [hep-ph]}
  \BibitemShut {NoStop}%
\bibitem [{\citenamefont {Roberts}\ \emph {et~al.}(2021)\citenamefont
  {Roberts}, \citenamefont {Richards}, \citenamefont {Horn},\ and\
  \citenamefont {Chang}}]{Roberts:2021nhw}%
  \BibitemOpen
  \bibfield  {author} {\bibinfo {author} {\bibfnamefont {C.~D.}\ \bibnamefont
  {Roberts}}, \bibinfo {author} {\bibfnamefont {D.~G.}\ \bibnamefont
  {Richards}}, \bibinfo {author} {\bibfnamefont {T.}~\bibnamefont {Horn}},\
  and\ \bibinfo {author} {\bibfnamefont {L.}~\bibnamefont {Chang}},\ }\bibfield
   {title} {\bibinfo {title} {{Insights into the emergence of mass from studies
  of pion and kaon structure}},\ }\href
  {https://doi.org/10.1016/j.ppnp.2021.103883} {\bibfield  {journal} {\bibinfo
  {journal} {Prog. Part. Nucl. Phys.}\ }\textbf {\bibinfo {volume} {120}},\
  \bibinfo {pages} {103883} (\bibinfo {year} {2021})},\ \Eprint
  {https://arxiv.org/abs/2102.01765} {arXiv:2102.01765 [hep-ph]} \BibitemShut
  {NoStop}%
\bibitem [{\citenamefont {Ding}\ \emph {et~al.}(2023)\citenamefont {Ding},
  \citenamefont {Roberts},\ and\ \citenamefont {Schmidt}}]{Ding:2022ows}%
  \BibitemOpen
  \bibfield  {author} {\bibinfo {author} {\bibfnamefont {M.}~\bibnamefont
  {Ding}}, \bibinfo {author} {\bibfnamefont {C.~D.}\ \bibnamefont {Roberts}},\
  and\ \bibinfo {author} {\bibfnamefont {S.~M.}\ \bibnamefont {Schmidt}},\
  }\bibfield  {title} {\bibinfo {title} {{Emergence of Hadron Mass and
  Structure}},\ }\href {https://doi.org/10.3390/particles6010004} {\bibfield
  {journal} {\bibinfo  {journal} {Particles}\ }\textbf {\bibinfo {volume}
  {6}},\ \bibinfo {pages} {57} (\bibinfo {year} {2023})},\ \Eprint
  {https://arxiv.org/abs/2211.07763} {arXiv:2211.07763 [hep-ph]} \BibitemShut
  {NoStop}%
\bibitem [{\citenamefont {Raya}\ \emph {et~al.}(2024)\citenamefont {Raya},
  \citenamefont {Bashir}, \citenamefont {Binosi}, \citenamefont {Roberts},\
  and\ \citenamefont {Rodr{\'\i}guez-Quintero}}]{Raya:2024ejx}%
  \BibitemOpen
  \bibfield  {author} {\bibinfo {author} {\bibfnamefont {K.}~\bibnamefont
  {Raya}}, \bibinfo {author} {\bibfnamefont {A.}~\bibnamefont {Bashir}},
  \bibinfo {author} {\bibfnamefont {D.}~\bibnamefont {Binosi}}, \bibinfo
  {author} {\bibfnamefont {C.~D.}\ \bibnamefont {Roberts}},\ and\ \bibinfo
  {author} {\bibfnamefont {J.}~\bibnamefont {Rodr{\'\i}guez-Quintero}},\
  }\bibfield  {title} {\bibinfo {title} {{Pseudoscalar Mesons and Emergent
  Mass}},\ }\href {https://doi.org/10.1007/s00601-024-01924-2} {\bibfield
  {journal} {\bibinfo  {journal} {Few Body Syst.}\ }\textbf {\bibinfo {volume}
  {65}},\ \bibinfo {pages} {60} (\bibinfo {year} {2024})},\ \Eprint
  {https://arxiv.org/abs/2403.00629} {arXiv:2403.00629 [hep-ph]} \BibitemShut
  {NoStop}%
\bibitem [{\citenamefont {Seth}\ \emph {et~al.}(2014)\citenamefont {Seth},
  \citenamefont {Dobbs}, \citenamefont {Tomaradze}, \citenamefont {Xiao},\ and\
  \citenamefont {Bonvicini}}]{Seth:2013eaa}%
  \BibitemOpen
  \bibfield  {author} {\bibinfo {author} {\bibfnamefont {K.~K.}\ \bibnamefont
  {Seth}}, \bibinfo {author} {\bibfnamefont {S.}~\bibnamefont {Dobbs}},
  \bibinfo {author} {\bibfnamefont {A.}~\bibnamefont {Tomaradze}}, \bibinfo
  {author} {\bibfnamefont {T.}~\bibnamefont {Xiao}},\ and\ \bibinfo {author}
  {\bibfnamefont {G.}~\bibnamefont {Bonvicini}},\ }\bibfield  {title} {\bibinfo
  {title} {{First Measurement of the Electromagnetic Form Factor of the Neutral
  Kaon at a Large Momentum Transfer and the Effect of $SU(3)$ Breaking}},\
  }\href {https://doi.org/10.1016/j.physletb.2014.02.003} {\bibfield  {journal}
  {\bibinfo  {journal} {Phys. Lett. B}\ }\textbf {\bibinfo {volume} {730}},\
  \bibinfo {pages} {332} (\bibinfo {year} {2014})},\ \Eprint
  {https://arxiv.org/abs/1307.6587} {arXiv:1307.6587 [hep-ex]} \BibitemShut
  {NoStop}%
\bibitem [{\citenamefont {Ablikim}\ \emph {et~al.}(2024)\citenamefont {Ablikim}
  \emph {et~al.}}]{BESIII:2023zsk}%
  \BibitemOpen
  \bibfield  {author} {\bibinfo {author} {\bibfnamefont {M.}~\bibnamefont
  {Ablikim}} \emph {et~al.} (\bibinfo {collaboration} {BESIII}),\ }\bibfield
  {title} {\bibinfo {title} {{Observation of significant flavor-SU(3) breaking
  in the kaon wave function at $12~{\rm GeV}^2<Q^2<25~{\rm GeV}^2$ and
  discovery of the charmless decay $\psi(3770)\to K_S^0K_L^0$}},\ }\href
  {https://doi.org/10.1103/PhysRevLett.132.131901} {\bibfield  {journal}
  {\bibinfo  {journal} {Phys. Rev. Lett.}\ }\textbf {\bibinfo {volume} {132}},\
  \bibinfo {pages} {131901} (\bibinfo {year} {2024})},\ \Eprint
  {https://arxiv.org/abs/2312.10962} {arXiv:2312.10962 [hep-ex]} \BibitemShut
  {NoStop}%
\bibitem [{\citenamefont {Efremov}\ and\ \citenamefont
  {Radyushkin}(1980)}]{Efremov:1979qk}%
  \BibitemOpen
  \bibfield  {author} {\bibinfo {author} {\bibfnamefont {A.~V.}\ \bibnamefont
  {Efremov}}\ and\ \bibinfo {author} {\bibfnamefont {A.~V.}\ \bibnamefont
  {Radyushkin}},\ }\bibfield  {title} {\bibinfo {title} {{Factorization and
  Asymptotical Behavior of Pion Form-Factor in QCD}},\ }\href
  {https://doi.org/10.1016/0370-2693(80)90869-2} {\bibfield  {journal}
  {\bibinfo  {journal} {Phys. Lett. B}\ }\textbf {\bibinfo {volume} {94}},\
  \bibinfo {pages} {245} (\bibinfo {year} {1980})}\BibitemShut {NoStop}%
\bibitem [{\citenamefont {Farrar}\ and\ \citenamefont
  {Jackson}(1979)}]{Farrar:1979aw}%
  \BibitemOpen
  \bibfield  {author} {\bibinfo {author} {\bibfnamefont {G.~R.}\ \bibnamefont
  {Farrar}}\ and\ \bibinfo {author} {\bibfnamefont {D.~R.}\ \bibnamefont
  {Jackson}},\ }\bibfield  {title} {\bibinfo {title} {{The Pion Form-Factor}},\
  }\href {https://doi.org/10.1103/PhysRevLett.43.246} {\bibfield  {journal}
  {\bibinfo  {journal} {Phys. Rev. Lett.}\ }\textbf {\bibinfo {volume} {43}},\
  \bibinfo {pages} {246} (\bibinfo {year} {1979})}\BibitemShut {NoStop}%
\bibitem [{\citenamefont {Lepage}\ and\ \citenamefont
  {Brodsky}(1979)}]{Lepage:1979zb}%
  \BibitemOpen
  \bibfield  {author} {\bibinfo {author} {\bibfnamefont {G.~P.}\ \bibnamefont
  {Lepage}}\ and\ \bibinfo {author} {\bibfnamefont {S.~J.}\ \bibnamefont
  {Brodsky}},\ }\bibfield  {title} {\bibinfo {title} {{Exclusive Processes in
  Quantum Chromodynamics: Evolution Equations for Hadronic Wave Functions and
  the Form-Factors of Mesons}},\ }\href
  {https://doi.org/10.1016/0370-2693(79)90554-9} {\bibfield  {journal}
  {\bibinfo  {journal} {Phys. Lett. B}\ }\textbf {\bibinfo {volume} {87}},\
  \bibinfo {pages} {359} (\bibinfo {year} {1979})}\BibitemShut {NoStop}%
\bibitem [{\citenamefont {Lepage}\ and\ \citenamefont
  {Brodsky}(1980)}]{Lepage:1980fj}%
  \BibitemOpen
  \bibfield  {author} {\bibinfo {author} {\bibfnamefont {G.~P.}\ \bibnamefont
  {Lepage}}\ and\ \bibinfo {author} {\bibfnamefont {S.~J.}\ \bibnamefont
  {Brodsky}},\ }\bibfield  {title} {\bibinfo {title} {{Exclusive Processes in
  Perturbative Quantum Chromodynamics}},\ }\href
  {https://doi.org/10.1103/PhysRevD.22.2157} {\bibfield  {journal} {\bibinfo
  {journal} {Phys. Rev. D}\ }\textbf {\bibinfo {volume} {22}},\ \bibinfo
  {pages} {2157} (\bibinfo {year} {1980})}\BibitemShut {NoStop}%
\bibitem [{\citenamefont {Gao}\ \emph {et~al.}(2017)\citenamefont {Gao},
  \citenamefont {Chang}, \citenamefont {Liu}, \citenamefont {Roberts},\ and\
  \citenamefont {Tandy}}]{Gao:2017mmp}%
  \BibitemOpen
  \bibfield  {author} {\bibinfo {author} {\bibfnamefont {F.}~\bibnamefont
  {Gao}}, \bibinfo {author} {\bibfnamefont {L.}~\bibnamefont {Chang}}, \bibinfo
  {author} {\bibfnamefont {Y.-X.}\ \bibnamefont {Liu}}, \bibinfo {author}
  {\bibfnamefont {C.~D.}\ \bibnamefont {Roberts}},\ and\ \bibinfo {author}
  {\bibfnamefont {P.~C.}\ \bibnamefont {Tandy}},\ }\bibfield  {title} {\bibinfo
  {title} {{Exposing strangeness: projections for kaon electromagnetic form
  factors}},\ }\href {https://doi.org/10.1103/PhysRevD.96.034024} {\bibfield
  {journal} {\bibinfo  {journal} {Phys. Rev. D}\ }\textbf {\bibinfo {volume}
  {96}},\ \bibinfo {pages} {034024} (\bibinfo {year} {2017})},\ \Eprint
  {https://arxiv.org/abs/1703.04875} {arXiv:1703.04875 [nucl-th]} \BibitemShut
  {NoStop}%
\bibitem [{\citenamefont {Hua}\ \emph {et~al.}(2022)\citenamefont {Hua} \emph
  {et~al.}}]{LatticeParton:2022zqc}%
  \BibitemOpen
  \bibfield  {author} {\bibinfo {author} {\bibfnamefont {J.}~\bibnamefont
  {Hua}} \emph {et~al.} (\bibinfo {collaboration} {Lattice Parton}),\
  }\bibfield  {title} {\bibinfo {title} {{Pion and Kaon Distribution Amplitudes
  from Lattice QCD}},\ }\href {https://doi.org/10.1103/PhysRevLett.129.132001}
  {\bibfield  {journal} {\bibinfo  {journal} {Phys. Rev. Lett.}\ }\textbf
  {\bibinfo {volume} {129}},\ \bibinfo {pages} {132001} (\bibinfo {year}
  {2022})},\ \Eprint {https://arxiv.org/abs/2201.09173} {arXiv:2201.09173
  [hep-lat]} \BibitemShut {NoStop}%
\bibitem [{\citenamefont {Chen}\ \emph {et~al.}(2025)\citenamefont {Chen},
  \citenamefont {Chen}, \citenamefont {Feng},\ and\ \citenamefont
  {Jia}}]{Chen:2024oem}%
  \BibitemOpen
  \bibfield  {author} {\bibinfo {author} {\bibfnamefont {L.-B.}\ \bibnamefont
  {Chen}}, \bibinfo {author} {\bibfnamefont {W.}~\bibnamefont {Chen}}, \bibinfo
  {author} {\bibfnamefont {F.}~\bibnamefont {Feng}},\ and\ \bibinfo {author}
  {\bibfnamefont {Y.}~\bibnamefont {Jia}},\ }\bibfield  {title} {\bibinfo
  {title} {{Confronting perturbative QCD with kaon electromagnetic form
  factors}},\ }\href {https://doi.org/10.1016/j.physletb.2025.139886}
  {\bibfield  {journal} {\bibinfo  {journal} {Phys. Lett. B}\ }\textbf
  {\bibinfo {volume} {870}},\ \bibinfo {pages} {139886} (\bibinfo {year}
  {2025})},\ \Eprint {https://arxiv.org/abs/2407.21120} {arXiv:2407.21120
  [hep-ph]} \BibitemShut {NoStop}%
\bibitem [{\citenamefont {Chai}\ and\ \citenamefont
  {Cheng}(2025)}]{Chai:2025xuz}%
  \BibitemOpen
  \bibfield  {author} {\bibinfo {author} {\bibfnamefont {J.}~\bibnamefont
  {Chai}}\ and\ \bibinfo {author} {\bibfnamefont {S.}~\bibnamefont {Cheng}},\
  }\bibfield  {title} {\bibinfo {title} {{Form factors of light pseudoscalar
  mesons from the perturbative QCD approach}},\ }\href
  {https://doi.org/10.1007/JHEP06(2025)229} {\bibfield  {journal} {\bibinfo
  {journal} {JHEP}\ }\textbf {\bibinfo {volume} {06}},\ \bibinfo {pages}
  {229}},\ \Eprint {https://arxiv.org/abs/2501.08783} {arXiv:2501.08783
  [hep-ph]} \BibitemShut {NoStop}%
\bibitem [{\citenamefont {Ding}\ \emph {et~al.}(2024)\citenamefont {Ding},
  \citenamefont {Gao}, \citenamefont {Hanlon}, \citenamefont {Mukherjee},
  \citenamefont {Petreczky}, \citenamefont {Shi}, \citenamefont {Syritsyn},
  \citenamefont {Zhang},\ and\ \citenamefont {Zhao}}]{Ding:2024lfj}%
  \BibitemOpen
  \bibfield  {author} {\bibinfo {author} {\bibfnamefont {H.-T.}\ \bibnamefont
  {Ding}}, \bibinfo {author} {\bibfnamefont {X.}~\bibnamefont {Gao}}, \bibinfo
  {author} {\bibfnamefont {A.~D.}\ \bibnamefont {Hanlon}}, \bibinfo {author}
  {\bibfnamefont {S.}~\bibnamefont {Mukherjee}}, \bibinfo {author}
  {\bibfnamefont {P.}~\bibnamefont {Petreczky}}, \bibinfo {author}
  {\bibfnamefont {Q.}~\bibnamefont {Shi}}, \bibinfo {author} {\bibfnamefont
  {S.}~\bibnamefont {Syritsyn}}, \bibinfo {author} {\bibfnamefont
  {R.}~\bibnamefont {Zhang}},\ and\ \bibinfo {author} {\bibfnamefont
  {Y.}~\bibnamefont {Zhao}},\ }\bibfield  {title} {\bibinfo {title} {{QCD
  Predictions for Meson Electromagnetic Form Factors at High Momenta: Testing
  Factorization in Exclusive Processes}},\ }\href
  {https://doi.org/10.1103/PhysRevLett.133.181902} {\bibfield  {journal}
  {\bibinfo  {journal} {Phys. Rev. Lett.}\ }\textbf {\bibinfo {volume} {133}},\
  \bibinfo {pages} {181902} (\bibinfo {year} {2024})},\ \Eprint
  {https://arxiv.org/abs/2404.04412} {arXiv:2404.04412 [hep-lat]} \BibitemShut
  {NoStop}%
\bibitem [{\citenamefont {Chang}\ \emph {et~al.}(2025)\citenamefont {Chang},
  \citenamefont {Liu}, \citenamefont {Raya},\ and\ \citenamefont
  {Sultan}}]{Chang:2025lrc}%
  \BibitemOpen
  \bibfield  {author} {\bibinfo {author} {\bibfnamefont {L.}~\bibnamefont
  {Chang}}, \bibinfo {author} {\bibfnamefont {Y.-B.}\ \bibnamefont {Liu}},
  \bibinfo {author} {\bibfnamefont {K.}~\bibnamefont {Raya}},\ and\ \bibinfo
  {author} {\bibfnamefont {M.~A.}\ \bibnamefont {Sultan}},\ }\bibfield  {title}
  {\bibinfo {title} {{Empirical determination of the kaon distribution
  amplitude}},\ }\href {https://doi.org/10.1103/3hd9-7wnt} {\bibfield
  {journal} {\bibinfo  {journal} {Phys. Rev. D}\ }\textbf {\bibinfo {volume}
  {112}},\ \bibinfo {pages} {114050} (\bibinfo {year} {2025})},\ \Eprint
  {https://arxiv.org/abs/2504.07372} {arXiv:2504.07372 [hep-ph]} \BibitemShut
  {NoStop}%
\bibitem [{\citenamefont {Shi}\ \emph {et~al.}(2014)\citenamefont {Shi},
  \citenamefont {Wang}, \citenamefont {Jiang}, \citenamefont {Cui},\ and\
  \citenamefont {Zong}}]{Shi:2014zpa}%
  \BibitemOpen
  \bibfield  {author} {\bibinfo {author} {\bibfnamefont {C.}~\bibnamefont
  {Shi}}, \bibinfo {author} {\bibfnamefont {Y.-L.}\ \bibnamefont {Wang}},
  \bibinfo {author} {\bibfnamefont {Y.}~\bibnamefont {Jiang}}, \bibinfo
  {author} {\bibfnamefont {Z.-F.}\ \bibnamefont {Cui}},\ and\ \bibinfo {author}
  {\bibfnamefont {H.-S.}\ \bibnamefont {Zong}},\ }\bibfield  {title} {\bibinfo
  {title} {{Locate QCD Critical End Point in a Continuum Model Study}},\ }\href
  {https://doi.org/10.1007/JHEP07(2014)014} {\bibfield  {journal} {\bibinfo
  {journal} {JHEP}\ }\textbf {\bibinfo {volume} {07}},\ \bibinfo {pages}
  {014}},\ \Eprint {https://arxiv.org/abs/1403.3797} {arXiv:1403.3797 [hep-ph]}
  \BibitemShut {NoStop}%
\bibitem [{\citenamefont {Cui}\ \emph {et~al.}(2020)\citenamefont {Cui},
  \citenamefont {Ding}, \citenamefont {Gao}, \citenamefont {Raya},
  \citenamefont {Binosi}, \citenamefont {Chang}, \citenamefont {Roberts},
  \citenamefont {Rodr{\'\i}guez-Quintero},\ and\ \citenamefont
  {Schmidt}}]{Cui:2020tdf}%
  \BibitemOpen
  \bibfield  {author} {\bibinfo {author} {\bibfnamefont {Z.-F.}\ \bibnamefont
  {Cui}}, \bibinfo {author} {\bibfnamefont {M.}~\bibnamefont {Ding}}, \bibinfo
  {author} {\bibfnamefont {F.}~\bibnamefont {Gao}}, \bibinfo {author}
  {\bibfnamefont {K.}~\bibnamefont {Raya}}, \bibinfo {author} {\bibfnamefont
  {D.}~\bibnamefont {Binosi}}, \bibinfo {author} {\bibfnamefont
  {L.}~\bibnamefont {Chang}}, \bibinfo {author} {\bibfnamefont {C.~D.}\
  \bibnamefont {Roberts}}, \bibinfo {author} {\bibfnamefont {J.}~\bibnamefont
  {Rodr{\'\i}guez-Quintero}},\ and\ \bibinfo {author} {\bibfnamefont {S.~M.}\
  \bibnamefont {Schmidt}},\ }\bibfield  {title} {\bibinfo {title} {{Kaon and
  pion parton distributions}},\ }\href
  {https://doi.org/10.1140/epjc/s10052-020-08578-4} {\bibfield  {journal}
  {\bibinfo  {journal} {Eur. Phys. J. C}\ }\textbf {\bibinfo {volume} {80}},\
  \bibinfo {pages} {1064} (\bibinfo {year} {2020})}\BibitemShut {NoStop}%
\bibitem [{\citenamefont {Swarnkar}\ and\ \citenamefont
  {Chakrabarti}(2015)}]{Swarnkar:2015osa}%
  \BibitemOpen
  \bibfield  {author} {\bibinfo {author} {\bibfnamefont {R.}~\bibnamefont
  {Swarnkar}}\ and\ \bibinfo {author} {\bibfnamefont {D.}~\bibnamefont
  {Chakrabarti}},\ }\bibfield  {title} {\bibinfo {title} {{Meson structure in
  light-front holographic QCD}},\ }\href
  {https://doi.org/10.1103/PhysRevD.92.074023} {\bibfield  {journal} {\bibinfo
  {journal} {Phys. Rev. D}\ }\textbf {\bibinfo {volume} {92}},\ \bibinfo
  {pages} {074023} (\bibinfo {year} {2015})},\ \Eprint
  {https://arxiv.org/abs/1507.01568} {arXiv:1507.01568 [hep-ph]} \BibitemShut
  {NoStop}%
\bibitem [{\citenamefont {Chang}\ \emph {et~al.}(2017)\citenamefont {Chang},
  \citenamefont {Brodsky},\ and\ \citenamefont {Li}}]{Chang:2016ouf}%
  \BibitemOpen
  \bibfield  {author} {\bibinfo {author} {\bibfnamefont {Q.}~\bibnamefont
  {Chang}}, \bibinfo {author} {\bibfnamefont {S.~J.}\ \bibnamefont {Brodsky}},\
  and\ \bibinfo {author} {\bibfnamefont {X.-Q.}\ \bibnamefont {Li}},\
  }\bibfield  {title} {\bibinfo {title} {{Light-front holographic distribution
  amplitudes of pseudoscalar mesons and their application to $B$-meson
  decays}},\ }\href {https://doi.org/10.1103/PhysRevD.95.094025} {\bibfield
  {journal} {\bibinfo  {journal} {Phys. Rev. D}\ }\textbf {\bibinfo {volume}
  {95}},\ \bibinfo {pages} {094025} (\bibinfo {year} {2017})},\ \Eprint
  {https://arxiv.org/abs/1612.05298} {arXiv:1612.05298 [hep-ph]} \BibitemShut
  {NoStop}%
\bibitem [{\citenamefont {Zhang}\ \emph {et~al.}(2020)\citenamefont {Zhang},
  \citenamefont {Honkala}, \citenamefont {Lin},\ and\ \citenamefont
  {Chen}}]{Zhang:2020gaj}%
  \BibitemOpen
  \bibfield  {author} {\bibinfo {author} {\bibfnamefont {R.}~\bibnamefont
  {Zhang}}, \bibinfo {author} {\bibfnamefont {C.}~\bibnamefont {Honkala}},
  \bibinfo {author} {\bibfnamefont {H.-W.}\ \bibnamefont {Lin}},\ and\ \bibinfo
  {author} {\bibfnamefont {J.-W.}\ \bibnamefont {Chen}},\ }\bibfield  {title}
  {\bibinfo {title} {{Pion and kaon distribution amplitudes in the continuum
  limit}},\ }\href {https://doi.org/10.1103/PhysRevD.102.094519} {\bibfield
  {journal} {\bibinfo  {journal} {Phys. Rev. D}\ }\textbf {\bibinfo {volume}
  {102}},\ \bibinfo {pages} {094519} (\bibinfo {year} {2020})},\ \Eprint
  {https://arxiv.org/abs/2005.13955} {arXiv:2005.13955 [hep-lat]} \BibitemShut
  {NoStop}%
\bibitem [{\citenamefont {Bali}\ \emph {et~al.}(2019)\citenamefont {Bali},
  \citenamefont {Braun}, \citenamefont {B\"urger}, \citenamefont {G\"ockeler},
  \citenamefont {Gruber}, \citenamefont {Hutzler}, \citenamefont {Korcyl},
  \citenamefont {Sch\"afer}, \citenamefont {Sternbeck},\ and\ \citenamefont
  {Wein}}]{RQCD:2019osh}%
  \BibitemOpen
  \bibfield  {author} {\bibinfo {author} {\bibfnamefont {G.~S.}\ \bibnamefont
  {Bali}}, \bibinfo {author} {\bibfnamefont {V.~M.}\ \bibnamefont {Braun}},
  \bibinfo {author} {\bibfnamefont {S.}~\bibnamefont {B\"urger}}, \bibinfo
  {author} {\bibfnamefont {M.}~\bibnamefont {G\"ockeler}}, \bibinfo {author}
  {\bibfnamefont {M.}~\bibnamefont {Gruber}}, \bibinfo {author} {\bibfnamefont
  {F.}~\bibnamefont {Hutzler}}, \bibinfo {author} {\bibfnamefont
  {P.}~\bibnamefont {Korcyl}}, \bibinfo {author} {\bibfnamefont
  {A.}~\bibnamefont {Sch\"afer}}, \bibinfo {author} {\bibfnamefont
  {A.}~\bibnamefont {Sternbeck}},\ and\ \bibinfo {author} {\bibfnamefont
  {P.}~\bibnamefont {Wein}} (\bibinfo {collaboration} {RQCD}),\ }\bibfield
  {title} {\bibinfo {title} {{Light-cone distribution amplitudes of
  pseudoscalar mesons from lattice QCD}},\ }\href
  {https://doi.org/10.1007/JHEP08(2019)065} {\bibfield  {journal} {\bibinfo
  {journal} {JHEP}\ }\textbf {\bibinfo {volume} {08}},\ \bibinfo {pages}
  {065}},\ \bibinfo {note} {[Addendum: JHEP 11, 037 (2020)]},\ \Eprint
  {https://arxiv.org/abs/1903.08038} {arXiv:1903.08038 [hep-lat]} \BibitemShut
  {NoStop}%
\bibitem [{\citenamefont {Mitter}\ \emph {et~al.}(2015)\citenamefont {Mitter},
  \citenamefont {Pawlowski},\ and\ \citenamefont
  {Strodthoff}}]{Mitter:2014wpa}%
  \BibitemOpen
  \bibfield  {author} {\bibinfo {author} {\bibfnamefont {M.}~\bibnamefont
  {Mitter}}, \bibinfo {author} {\bibfnamefont {J.~M.}\ \bibnamefont
  {Pawlowski}},\ and\ \bibinfo {author} {\bibfnamefont {N.}~\bibnamefont
  {Strodthoff}},\ }\bibfield  {title} {\bibinfo {title} {{Chiral symmetry
  breaking in continuum QCD}},\ }\href
  {https://doi.org/10.1103/PhysRevD.91.054035} {\bibfield  {journal} {\bibinfo
  {journal} {Phys. Rev.}\ }\textbf {\bibinfo {volume} {D91}},\ \bibinfo {pages}
  {054035} (\bibinfo {year} {2015})},\ \Eprint
  {https://arxiv.org/abs/1411.7978} {arXiv:1411.7978 [hep-ph]} \BibitemShut
  {NoStop}%
\bibitem [{\citenamefont {Braun}\ \emph {et~al.}(2016)\citenamefont {Braun},
  \citenamefont {Fister}, \citenamefont {Pawlowski},\ and\ \citenamefont
  {Rennecke}}]{Braun:2014ata}%
  \BibitemOpen
  \bibfield  {author} {\bibinfo {author} {\bibfnamefont {J.}~\bibnamefont
  {Braun}}, \bibinfo {author} {\bibfnamefont {L.}~\bibnamefont {Fister}},
  \bibinfo {author} {\bibfnamefont {J.~M.}\ \bibnamefont {Pawlowski}},\ and\
  \bibinfo {author} {\bibfnamefont {F.}~\bibnamefont {Rennecke}},\ }\bibfield
  {title} {\bibinfo {title} {{From Quarks and Gluons to Hadrons: Chiral
  Symmetry Breaking in Dynamical QCD}},\ }\href
  {https://doi.org/10.1103/PhysRevD.94.034016} {\bibfield  {journal} {\bibinfo
  {journal} {Phys. Rev.}\ }\textbf {\bibinfo {volume} {D94}},\ \bibinfo {pages}
  {034016} (\bibinfo {year} {2016})},\ \Eprint
  {https://arxiv.org/abs/1412.1045} {arXiv:1412.1045 [hep-ph]} \BibitemShut
  {NoStop}%
\bibitem [{\citenamefont {Rennecke}(2015)}]{Rennecke:2015eba}%
  \BibitemOpen
  \bibfield  {author} {\bibinfo {author} {\bibfnamefont {F.}~\bibnamefont
  {Rennecke}},\ }\bibfield  {title} {\bibinfo {title} {{Vacuum structure of
  vector mesons in QCD}},\ }\href {https://doi.org/10.1103/PhysRevD.92.076012}
  {\bibfield  {journal} {\bibinfo  {journal} {Phys. Rev.}\ }\textbf {\bibinfo
  {volume} {D92}},\ \bibinfo {pages} {076012} (\bibinfo {year} {2015})},\
  \Eprint {https://arxiv.org/abs/1504.03585} {arXiv:1504.03585 [hep-ph]}
  \BibitemShut {NoStop}%
\bibitem [{\citenamefont {Cyrol}\ \emph {et~al.}(2016)\citenamefont {Cyrol},
  \citenamefont {Fister}, \citenamefont {Mitter}, \citenamefont {Pawlowski},\
  and\ \citenamefont {Strodthoff}}]{Cyrol:2016tym}%
  \BibitemOpen
  \bibfield  {author} {\bibinfo {author} {\bibfnamefont {A.~K.}\ \bibnamefont
  {Cyrol}}, \bibinfo {author} {\bibfnamefont {L.}~\bibnamefont {Fister}},
  \bibinfo {author} {\bibfnamefont {M.}~\bibnamefont {Mitter}}, \bibinfo
  {author} {\bibfnamefont {J.~M.}\ \bibnamefont {Pawlowski}},\ and\ \bibinfo
  {author} {\bibfnamefont {N.}~\bibnamefont {Strodthoff}},\ }\bibfield  {title}
  {\bibinfo {title} {{Landau gauge Yang-Mills correlation functions}},\ }\href
  {https://doi.org/10.1103/PhysRevD.94.054005} {\bibfield  {journal} {\bibinfo
  {journal} {Phys. Rev.}\ }\textbf {\bibinfo {volume} {D94}},\ \bibinfo {pages}
  {054005} (\bibinfo {year} {2016})},\ \Eprint
  {https://arxiv.org/abs/1605.01856} {arXiv:1605.01856 [hep-ph]} \BibitemShut
  {NoStop}%
\bibitem [{\citenamefont {Cyrol}\ \emph {et~al.}(2018)\citenamefont {Cyrol},
  \citenamefont {Mitter}, \citenamefont {Pawlowski},\ and\ \citenamefont
  {Strodthoff}}]{Cyrol:2017ewj}%
  \BibitemOpen
  \bibfield  {author} {\bibinfo {author} {\bibfnamefont {A.~K.}\ \bibnamefont
  {Cyrol}}, \bibinfo {author} {\bibfnamefont {M.}~\bibnamefont {Mitter}},
  \bibinfo {author} {\bibfnamefont {J.~M.}\ \bibnamefont {Pawlowski}},\ and\
  \bibinfo {author} {\bibfnamefont {N.}~\bibnamefont {Strodthoff}},\ }\bibfield
   {title} {\bibinfo {title} {{Nonperturbative quark, gluon, and meson
  correlators of unquenched QCD}},\ }\href
  {https://doi.org/10.1103/PhysRevD.97.054006} {\bibfield  {journal} {\bibinfo
  {journal} {Phys. Rev.}\ }\textbf {\bibinfo {volume} {D97}},\ \bibinfo {pages}
  {054006} (\bibinfo {year} {2018})},\ \Eprint
  {https://arxiv.org/abs/1706.06326} {arXiv:1706.06326 [hep-ph]} \BibitemShut
  {NoStop}%
\bibitem [{\citenamefont {Corell}\ \emph {et~al.}(2018)\citenamefont {Corell},
  \citenamefont {Cyrol}, \citenamefont {Mitter}, \citenamefont {Pawlowski},\
  and\ \citenamefont {Strodthoff}}]{Corell:2018yil}%
  \BibitemOpen
  \bibfield  {author} {\bibinfo {author} {\bibfnamefont {L.}~\bibnamefont
  {Corell}}, \bibinfo {author} {\bibfnamefont {A.~K.}\ \bibnamefont {Cyrol}},
  \bibinfo {author} {\bibfnamefont {M.}~\bibnamefont {Mitter}}, \bibinfo
  {author} {\bibfnamefont {J.~M.}\ \bibnamefont {Pawlowski}},\ and\ \bibinfo
  {author} {\bibfnamefont {N.}~\bibnamefont {Strodthoff}},\ }\bibfield  {title}
  {\bibinfo {title} {{Correlation functions of three-dimensional Yang-Mills
  theory from the FRG}},\ }\href {https://doi.org/10.21468/SciPostPhys.5.6.066}
  {\bibfield  {journal} {\bibinfo  {journal} {SciPost Phys.}\ }\textbf
  {\bibinfo {volume} {5}},\ \bibinfo {pages} {066} (\bibinfo {year} {2018})},\
  \Eprint {https://arxiv.org/abs/1803.10092} {arXiv:1803.10092 [hep-ph]}
  \BibitemShut {NoStop}%
\bibitem [{\citenamefont {Fu}\ \emph {et~al.}(2020)\citenamefont {Fu},
  \citenamefont {Pawlowski},\ and\ \citenamefont {Rennecke}}]{Fu:2019hdw}%
  \BibitemOpen
  \bibfield  {author} {\bibinfo {author} {\bibfnamefont {W.-j.}\ \bibnamefont
  {Fu}}, \bibinfo {author} {\bibfnamefont {J.~M.}\ \bibnamefont {Pawlowski}},\
  and\ \bibinfo {author} {\bibfnamefont {F.}~\bibnamefont {Rennecke}},\
  }\bibfield  {title} {\bibinfo {title} {{QCD phase structure at finite
  temperature and density}},\ }\href
  {https://doi.org/10.1103/PhysRevD.101.054032} {\bibfield  {journal} {\bibinfo
   {journal} {Phys. Rev. D}\ }\textbf {\bibinfo {volume} {101}},\ \bibinfo
  {pages} {054032} (\bibinfo {year} {2020})},\ \Eprint
  {https://arxiv.org/abs/1909.02991} {arXiv:1909.02991 [hep-ph]} \BibitemShut
  {NoStop}%
\bibitem [{\citenamefont {Ihssen}\ \emph {et~al.}(2024)\citenamefont {Ihssen},
  \citenamefont {Pawlowski}, \citenamefont {Sattler},\ and\ \citenamefont
  {Wink}}]{Ihssen:2024miv}%
  \BibitemOpen
  \bibfield  {author} {\bibinfo {author} {\bibfnamefont {F.}~\bibnamefont
  {Ihssen}}, \bibinfo {author} {\bibfnamefont {J.~M.}\ \bibnamefont
  {Pawlowski}}, \bibinfo {author} {\bibfnamefont {F.~R.}\ \bibnamefont
  {Sattler}},\ and\ \bibinfo {author} {\bibfnamefont {N.}~\bibnamefont
  {Wink}},\ }\bibfield  {title} {\bibinfo {title} {{Towards quantitative
  precision in functional QCD I}},\ }\href@noop {} {\  (\bibinfo {year}
  {2024})},\ \Eprint {https://arxiv.org/abs/2408.08413} {arXiv:2408.08413
  [hep-ph]} \BibitemShut {NoStop}%
\bibitem [{\citenamefont {Pawlowski}\ \emph {et~al.}(2025)\citenamefont
  {Pawlowski}, \citenamefont {Rennecke},\ and\ \citenamefont
  {Sattler}}]{Pawlowski:2025jpg}%
  \BibitemOpen
  \bibfield  {author} {\bibinfo {author} {\bibfnamefont {J.~M.}\ \bibnamefont
  {Pawlowski}}, \bibinfo {author} {\bibfnamefont {F.}~\bibnamefont
  {Rennecke}},\ and\ \bibinfo {author} {\bibfnamefont {F.~R.}\ \bibnamefont
  {Sattler}},\ }\bibfield  {title} {\bibinfo {title} {{Inhomogeneous
  instabilities in high-density QCD}},\ }\href@noop {} {\  (\bibinfo {year}
  {2025})},\ \Eprint {https://arxiv.org/abs/2512.20510} {arXiv:2512.20510
  [hep-ph]} \BibitemShut {NoStop}%
\bibitem [{\citenamefont {Fu}\ \emph {et~al.}(2026)\citenamefont {Fu},
  \citenamefont {Huang}, \citenamefont {Pawlowski}, \citenamefont {Rennecke},
  \citenamefont {Wen},\ and\ \citenamefont {Yin}}]{Fu:2026qnl}%
  \BibitemOpen
  \bibfield  {author} {\bibinfo {author} {\bibfnamefont {W.-j.}\ \bibnamefont
  {Fu}}, \bibinfo {author} {\bibfnamefont {C.}~\bibnamefont {Huang}}, \bibinfo
  {author} {\bibfnamefont {J.~M.}\ \bibnamefont {Pawlowski}}, \bibinfo {author}
  {\bibfnamefont {F.}~\bibnamefont {Rennecke}}, \bibinfo {author}
  {\bibfnamefont {R.}~\bibnamefont {Wen}},\ and\ \bibinfo {author}
  {\bibfnamefont {S.}~\bibnamefont {Yin}},\ }\bibfield  {title} {\bibinfo
  {title} {{Strangeness neutrality and the QCD phase diagram}},\ }\href@noop {}
  {\  (\bibinfo {year} {2026})},\ \Eprint {https://arxiv.org/abs/2603.13455}
  {arXiv:2603.13455 [hep-ph]} \BibitemShut {NoStop}%
\bibitem [{\citenamefont {Dupuis}\ \emph {et~al.}(2021)\citenamefont {Dupuis},
  \citenamefont {Canet}, \citenamefont {Eichhorn}, \citenamefont {Metzner},
  \citenamefont {Pawlowski}, \citenamefont {Tissier},\ and\ \citenamefont
  {Wschebor}}]{Dupuis:2020fhh}%
  \BibitemOpen
  \bibfield  {author} {\bibinfo {author} {\bibfnamefont {N.}~\bibnamefont
  {Dupuis}}, \bibinfo {author} {\bibfnamefont {L.}~\bibnamefont {Canet}},
  \bibinfo {author} {\bibfnamefont {A.}~\bibnamefont {Eichhorn}}, \bibinfo
  {author} {\bibfnamefont {W.}~\bibnamefont {Metzner}}, \bibinfo {author}
  {\bibfnamefont {J.~M.}\ \bibnamefont {Pawlowski}}, \bibinfo {author}
  {\bibfnamefont {M.}~\bibnamefont {Tissier}},\ and\ \bibinfo {author}
  {\bibfnamefont {N.}~\bibnamefont {Wschebor}},\ }\bibfield  {title} {\bibinfo
  {title} {{The nonperturbative functional renormalization group and its
  applications}},\ }\href {https://doi.org/10.1016/j.physrep.2021.01.001}
  {\bibfield  {journal} {\bibinfo  {journal} {Phys. Rept.}\ }\textbf {\bibinfo
  {volume} {910}},\ \bibinfo {pages} {1} (\bibinfo {year} {2021})},\ \Eprint
  {https://arxiv.org/abs/2006.04853} {arXiv:2006.04853 [cond-mat.stat-mech]}
  \BibitemShut {NoStop}%
\bibitem [{\citenamefont {Fu}(2022)}]{Fu:2022gou}%
  \BibitemOpen
  \bibfield  {author} {\bibinfo {author} {\bibfnamefont {W.-j.}\ \bibnamefont
  {Fu}},\ }\bibfield  {title} {\bibinfo {title} {{QCD at finite temperature and
  density within the fRG approach: an overview}},\ }\href
  {https://doi.org/10.1088/1572-9494/ac86be} {\bibfield  {journal} {\bibinfo
  {journal} {Commun. Theor. Phys.}\ }\textbf {\bibinfo {volume} {74}},\
  \bibinfo {pages} {097304} (\bibinfo {year} {2022})},\ \Eprint
  {https://arxiv.org/abs/2205.00468} {arXiv:2205.00468 [hep-ph]} \BibitemShut
  {NoStop}%
\bibitem [{\citenamefont {Rennecke}(2025)}]{Rennecke:2025bcw}%
  \BibitemOpen
  \bibfield  {author} {\bibinfo {author} {\bibfnamefont {F.}~\bibnamefont
  {Rennecke}},\ }\bibfield  {title} {\bibinfo {title} {{QCD phase structure
  {\&} equation of state: A functional perspective}},\ }in\ \href@noop {}
  {\emph {\bibinfo {booktitle} {{31st International Conference on
  Ultra-relativistic Nucleus-Nucleus Collisions}}}}\ (\bibinfo {year} {2025})\
  \Eprint {https://arxiv.org/abs/2510.11270} {arXiv:2510.11270 [hep-ph]}
  \BibitemShut {NoStop}%
\bibitem [{\citenamefont {Fischer}\ and\ \citenamefont
  {Pawlowski}(2026)}]{Fischer:2026uni}%
  \BibitemOpen
  \bibfield  {author} {\bibinfo {author} {\bibfnamefont {C.~S.}\ \bibnamefont
  {Fischer}}\ and\ \bibinfo {author} {\bibfnamefont {J.~M.}\ \bibnamefont
  {Pawlowski}},\ }\bibfield  {title} {\bibinfo {title} {{Phase structure and
  observables at high densities from first principles QCD}},\ }\href@noop {} {\
   (\bibinfo {year} {2026})},\ \Eprint {https://arxiv.org/abs/2603.11135}
  {arXiv:2603.11135 [hep-ph]} \BibitemShut {NoStop}%
\bibitem [{\citenamefont {Fu}\ \emph {et~al.}(2023)\citenamefont {Fu},
  \citenamefont {Huang}, \citenamefont {Pawlowski},\ and\ \citenamefont
  {Tan}}]{Fu:2022uow}%
  \BibitemOpen
  \bibfield  {author} {\bibinfo {author} {\bibfnamefont {W.-j.}\ \bibnamefont
  {Fu}}, \bibinfo {author} {\bibfnamefont {C.}~\bibnamefont {Huang}}, \bibinfo
  {author} {\bibfnamefont {J.~M.}\ \bibnamefont {Pawlowski}},\ and\ \bibinfo
  {author} {\bibfnamefont {Y.-y.}\ \bibnamefont {Tan}},\ }\bibfield  {title}
  {\bibinfo {title} {{Four-quark scatterings in QCD I}},\ }\href
  {https://doi.org/10.21468/SciPostPhys.14.4.069} {\bibfield  {journal}
  {\bibinfo  {journal} {SciPost Phys.}\ }\textbf {\bibinfo {volume} {14}},\
  \bibinfo {pages} {069} (\bibinfo {year} {2023})},\ \Eprint
  {https://arxiv.org/abs/2209.13120} {arXiv:2209.13120 [hep-ph]} \BibitemShut
  {NoStop}%
\bibitem [{\citenamefont {Fu}\ \emph {et~al.}(2024)\citenamefont {Fu},
  \citenamefont {Huang}, \citenamefont {Pawlowski},\ and\ \citenamefont
  {Tan}}]{Fu:2024ysj}%
  \BibitemOpen
  \bibfield  {author} {\bibinfo {author} {\bibfnamefont {W.-j.}\ \bibnamefont
  {Fu}}, \bibinfo {author} {\bibfnamefont {C.}~\bibnamefont {Huang}}, \bibinfo
  {author} {\bibfnamefont {J.~M.}\ \bibnamefont {Pawlowski}},\ and\ \bibinfo
  {author} {\bibfnamefont {Y.-y.}\ \bibnamefont {Tan}},\ }\bibfield  {title}
  {\bibinfo {title} {{Four-quark scatterings in QCD II}},\ }\href@noop {} {\
  (\bibinfo {year} {2024})},\ \Eprint {https://arxiv.org/abs/2401.07638}
  {arXiv:2401.07638 [hep-ph]} \BibitemShut {NoStop}%
\bibitem [{\citenamefont {Ji}(2013)}]{Ji:2013dva}%
  \BibitemOpen
  \bibfield  {author} {\bibinfo {author} {\bibfnamefont {X.}~\bibnamefont
  {Ji}},\ }\bibfield  {title} {\bibinfo {title} {{Parton Physics on a Euclidean
  Lattice}},\ }\href {https://doi.org/10.1103/PhysRevLett.110.262002}
  {\bibfield  {journal} {\bibinfo  {journal} {Phys. Rev. Lett.}\ }\textbf
  {\bibinfo {volume} {110}},\ \bibinfo {pages} {262002} (\bibinfo {year}
  {2013})},\ \Eprint {https://arxiv.org/abs/1305.1539} {arXiv:1305.1539
  [hep-ph]} \BibitemShut {NoStop}%
\bibitem [{\citenamefont {Ji}\ \emph {et~al.}(2021)\citenamefont {Ji},
  \citenamefont {Liu}, \citenamefont {Liu}, \citenamefont {Zhang},\ and\
  \citenamefont {Zhao}}]{Ji:2020ect}%
  \BibitemOpen
  \bibfield  {author} {\bibinfo {author} {\bibfnamefont {X.}~\bibnamefont
  {Ji}}, \bibinfo {author} {\bibfnamefont {Y.-S.}\ \bibnamefont {Liu}},
  \bibinfo {author} {\bibfnamefont {Y.}~\bibnamefont {Liu}}, \bibinfo {author}
  {\bibfnamefont {J.-H.}\ \bibnamefont {Zhang}},\ and\ \bibinfo {author}
  {\bibfnamefont {Y.}~\bibnamefont {Zhao}},\ }\bibfield  {title} {\bibinfo
  {title} {{Large-momentum effective theory}},\ }\href
  {https://doi.org/10.1103/RevModPhys.93.035005} {\bibfield  {journal}
  {\bibinfo  {journal} {Rev. Mod. Phys.}\ }\textbf {\bibinfo {volume} {93}},\
  \bibinfo {pages} {035005} (\bibinfo {year} {2021})},\ \Eprint
  {https://arxiv.org/abs/2004.03543} {arXiv:2004.03543 [hep-ph]} \BibitemShut
  {NoStop}%
\bibitem [{\citenamefont {Ji}\ \emph {et~al.}(2017)\citenamefont {Ji},
  \citenamefont {Zhang},\ and\ \citenamefont {Zhao}}]{Ji:2017rah}%
  \BibitemOpen
  \bibfield  {author} {\bibinfo {author} {\bibfnamefont {X.}~\bibnamefont
  {Ji}}, \bibinfo {author} {\bibfnamefont {J.-H.}\ \bibnamefont {Zhang}},\ and\
  \bibinfo {author} {\bibfnamefont {Y.}~\bibnamefont {Zhao}},\ }\bibfield
  {title} {\bibinfo {title} {{More On Large-Momentum Effective Theory Approach
  to Parton Physics}},\ }\href
  {https://doi.org/10.1016/j.nuclphysb.2017.09.001} {\bibfield  {journal}
  {\bibinfo  {journal} {Nucl. Phys. B}\ }\textbf {\bibinfo {volume} {924}},\
  \bibinfo {pages} {366} (\bibinfo {year} {2017})},\ \Eprint
  {https://arxiv.org/abs/1706.07416} {arXiv:1706.07416 [hep-ph]} \BibitemShut
  {NoStop}%
\bibitem [{\citenamefont {Braun}\ \emph {et~al.}(2010)\citenamefont {Braun},
  \citenamefont {Gies},\ and\ \citenamefont {Pawlowski}}]{Braun:2007bx}%
  \BibitemOpen
  \bibfield  {author} {\bibinfo {author} {\bibfnamefont {J.}~\bibnamefont
  {Braun}}, \bibinfo {author} {\bibfnamefont {H.}~\bibnamefont {Gies}},\ and\
  \bibinfo {author} {\bibfnamefont {J.~M.}\ \bibnamefont {Pawlowski}},\
  }\bibfield  {title} {\bibinfo {title} {{Quark Confinement from Color
  Confinement}},\ }\href {https://doi.org/10.1016/j.physletb.2010.01.009}
  {\bibfield  {journal} {\bibinfo  {journal} {Phys.Lett.}\ }\textbf {\bibinfo
  {volume} {B684}},\ \bibinfo {pages} {262} (\bibinfo {year} {2010})},\ \Eprint
  {https://arxiv.org/abs/0708.2413} {arXiv:0708.2413 [hep-th]} \BibitemShut
  {NoStop}%
\bibitem [{\citenamefont {Fischer}\ \emph {et~al.}(2009)\citenamefont
  {Fischer}, \citenamefont {Maas},\ and\ \citenamefont
  {Pawlowski}}]{Fischer:2008uz}%
  \BibitemOpen
  \bibfield  {author} {\bibinfo {author} {\bibfnamefont {C.~S.}\ \bibnamefont
  {Fischer}}, \bibinfo {author} {\bibfnamefont {A.}~\bibnamefont {Maas}},\ and\
  \bibinfo {author} {\bibfnamefont {J.~M.}\ \bibnamefont {Pawlowski}},\
  }\bibfield  {title} {\bibinfo {title} {{On the infrared behavior of Landau
  gauge Yang-Mills theory}},\ }\href
  {https://doi.org/10.1016/j.aop.2009.07.009} {\bibfield  {journal} {\bibinfo
  {journal} {Annals Phys.}\ }\textbf {\bibinfo {volume} {324}},\ \bibinfo
  {pages} {2408} (\bibinfo {year} {2009})},\ \Eprint
  {https://arxiv.org/abs/0810.1987} {arXiv:0810.1987 [hep-ph]} \BibitemShut
  {NoStop}%
\bibitem [{\citenamefont {Fister}\ and\ \citenamefont
  {Pawlowski}(2013)}]{Fister:2013bh}%
  \BibitemOpen
  \bibfield  {author} {\bibinfo {author} {\bibfnamefont {L.}~\bibnamefont
  {Fister}}\ and\ \bibinfo {author} {\bibfnamefont {J.~M.}\ \bibnamefont
  {Pawlowski}},\ }\bibfield  {title} {\bibinfo {title} {{Confinement from
  Correlation Functions}},\ }\href {https://doi.org/10.1103/PhysRevD.88.045010}
  {\bibfield  {journal} {\bibinfo  {journal} {Phys.Rev.}\ }\textbf {\bibinfo
  {volume} {D88}},\ \bibinfo {pages} {045010} (\bibinfo {year} {2013})},\
  \Eprint {https://arxiv.org/abs/1301.4163} {arXiv:1301.4163 [hep-ph]}
  \BibitemShut {NoStop}%
\bibitem [{\citenamefont {Williams}(2015)}]{Williams:2014iea}%
  \BibitemOpen
  \bibfield  {author} {\bibinfo {author} {\bibfnamefont {R.}~\bibnamefont
  {Williams}},\ }\bibfield  {title} {\bibinfo {title} {{The quark-gluon vertex
  in Landau gauge bound-state studies}},\ }\href
  {https://doi.org/10.1140/epja/i2015-15057-4} {\bibfield  {journal} {\bibinfo
  {journal} {Eur. Phys. J. A}\ }\textbf {\bibinfo {volume} {51}},\ \bibinfo
  {pages} {57} (\bibinfo {year} {2015})},\ \Eprint
  {https://arxiv.org/abs/1404.2545} {arXiv:1404.2545 [hep-ph]} \BibitemShut
  {NoStop}%
\bibitem [{\citenamefont {Williams}\ \emph {et~al.}(2016)\citenamefont
  {Williams}, \citenamefont {Fischer},\ and\ \citenamefont
  {Heupel}}]{Williams:2015cvx}%
  \BibitemOpen
  \bibfield  {author} {\bibinfo {author} {\bibfnamefont {R.}~\bibnamefont
  {Williams}}, \bibinfo {author} {\bibfnamefont {C.~S.}\ \bibnamefont
  {Fischer}},\ and\ \bibinfo {author} {\bibfnamefont {W.}~\bibnamefont
  {Heupel}},\ }\bibfield  {title} {\bibinfo {title} {{Light mesons in QCD and
  unquenching effects from the 3PI effective action}},\ }\href
  {https://doi.org/10.1103/PhysRevD.93.034026} {\bibfield  {journal} {\bibinfo
  {journal} {Phys. Rev.}\ }\textbf {\bibinfo {volume} {D93}},\ \bibinfo {pages}
  {034026} (\bibinfo {year} {2016})},\ \Eprint
  {https://arxiv.org/abs/1512.00455} {arXiv:1512.00455 [hep-ph]} \BibitemShut
  {NoStop}%
\bibitem [{\citenamefont {Aguilar}\ \emph {et~al.}(2017)\citenamefont
  {Aguilar}, \citenamefont {Cardona}, \citenamefont {Ferreira},\ and\
  \citenamefont {Papavassiliou}}]{Aguilar:2016lbe}%
  \BibitemOpen
  \bibfield  {author} {\bibinfo {author} {\bibfnamefont {A.~C.}\ \bibnamefont
  {Aguilar}}, \bibinfo {author} {\bibfnamefont {J.~C.}\ \bibnamefont
  {Cardona}}, \bibinfo {author} {\bibfnamefont {M.~N.}\ \bibnamefont
  {Ferreira}},\ and\ \bibinfo {author} {\bibfnamefont {J.}~\bibnamefont
  {Papavassiliou}},\ }\bibfield  {title} {\bibinfo {title} {{Non-Abelian
  Ball-Chiu vertex for arbitrary Euclidean momenta}},\ }\href
  {https://doi.org/10.1103/PhysRevD.96.014029} {\bibfield  {journal} {\bibinfo
  {journal} {Phys. Rev. D}\ }\textbf {\bibinfo {volume} {96}},\ \bibinfo
  {pages} {014029} (\bibinfo {year} {2017})},\ \Eprint
  {https://arxiv.org/abs/1610.06158} {arXiv:1610.06158 [hep-ph]} \BibitemShut
  {NoStop}%
\bibitem [{\citenamefont {Gao}\ \emph {et~al.}(2021)\citenamefont {Gao},
  \citenamefont {Papavassiliou},\ and\ \citenamefont
  {Pawlowski}}]{Gao:2021wun}%
  \BibitemOpen
  \bibfield  {author} {\bibinfo {author} {\bibfnamefont {F.}~\bibnamefont
  {Gao}}, \bibinfo {author} {\bibfnamefont {J.}~\bibnamefont {Papavassiliou}},\
  and\ \bibinfo {author} {\bibfnamefont {J.~M.}\ \bibnamefont {Pawlowski}},\
  }\bibfield  {title} {\bibinfo {title} {{Fully coupled functional equations
  for the quark sector of QCD}},\ }\href
  {https://doi.org/10.1103/PhysRevD.103.094013} {\bibfield  {journal} {\bibinfo
   {journal} {Phys. Rev. D}\ }\textbf {\bibinfo {volume} {103}},\ \bibinfo
  {pages} {094013} (\bibinfo {year} {2021})},\ \Eprint
  {https://arxiv.org/abs/2102.13053} {arXiv:2102.13053 [hep-ph]} \BibitemShut
  {NoStop}%
\bibitem [{\citenamefont {Chang}\ \emph {et~al.}(2021)\citenamefont {Chang},
  \citenamefont {Liu}, \citenamefont {Raya}, \citenamefont
  {Rodr\'\i{}guez-Quintero},\ and\ \citenamefont {Yang}}]{Chang:2021vvx}%
  \BibitemOpen
  \bibfield  {author} {\bibinfo {author} {\bibfnamefont {L.}~\bibnamefont
  {Chang}}, \bibinfo {author} {\bibfnamefont {Y.-B.}\ \bibnamefont {Liu}},
  \bibinfo {author} {\bibfnamefont {K.}~\bibnamefont {Raya}}, \bibinfo {author}
  {\bibfnamefont {J.}~\bibnamefont {Rodr\'\i{}guez-Quintero}},\ and\ \bibinfo
  {author} {\bibfnamefont {Y.-B.}\ \bibnamefont {Yang}},\ }\bibfield  {title}
  {\bibinfo {title} {{Linking continuum and lattice quark mass functions via an
  effective charge}},\ }\href {https://doi.org/10.1103/PhysRevD.104.094509}
  {\bibfield  {journal} {\bibinfo  {journal} {Phys. Rev. D}\ }\textbf {\bibinfo
  {volume} {104}},\ \bibinfo {pages} {094509} (\bibinfo {year} {2021})},\
  \Eprint {https://arxiv.org/abs/2105.06596} {arXiv:2105.06596 [hep-lat]}
  \BibitemShut {NoStop}%
\bibitem [{\citenamefont {Bowman}\ \emph {et~al.}(2005)\citenamefont {Bowman},
  \citenamefont {Heller}, \citenamefont {Leinweber}, \citenamefont
  {Parappilly}, \citenamefont {Williams},\ and\ \citenamefont
  {Zhang}}]{Bowman:2005vx}%
  \BibitemOpen
  \bibfield  {author} {\bibinfo {author} {\bibfnamefont {P.~O.}\ \bibnamefont
  {Bowman}}, \bibinfo {author} {\bibfnamefont {U.~M.}\ \bibnamefont {Heller}},
  \bibinfo {author} {\bibfnamefont {D.~B.}\ \bibnamefont {Leinweber}}, \bibinfo
  {author} {\bibfnamefont {M.~B.}\ \bibnamefont {Parappilly}}, \bibinfo
  {author} {\bibfnamefont {A.~G.}\ \bibnamefont {Williams}},\ and\ \bibinfo
  {author} {\bibfnamefont {J.-b.}\ \bibnamefont {Zhang}},\ }\bibfield  {title}
  {\bibinfo {title} {{Unquenched quark propagator in Landau gauge}},\ }\href
  {https://doi.org/10.1103/PhysRevD.71.054507} {\bibfield  {journal} {\bibinfo
  {journal} {Phys. Rev. D}\ }\textbf {\bibinfo {volume} {71}},\ \bibinfo
  {pages} {054507} (\bibinfo {year} {2005})},\ \Eprint
  {https://arxiv.org/abs/hep-lat/0501019} {arXiv:hep-lat/0501019} \BibitemShut
  {NoStop}%
\bibitem [{\citenamefont {Chang}\ \emph {et~al.}(2026)\citenamefont {Chang},
  \citenamefont {Fu}, \citenamefont {Huang}, \citenamefont {Pawlowski},\ and\
  \citenamefont {Tan}}]{PionPDA:2026}%
  \BibitemOpen
  \bibfield  {author} {\bibinfo {author} {\bibfnamefont {L.}~\bibnamefont
  {Chang}}, \bibinfo {author} {\bibfnamefont {W.-j.}\ \bibnamefont {Fu}},
  \bibinfo {author} {\bibfnamefont {C.}~\bibnamefont {Huang}}, \bibinfo
  {author} {\bibfnamefont {J.~M.}\ \bibnamefont {Pawlowski}},\ and\ \bibinfo
  {author} {\bibfnamefont {Y.-y.}\ \bibnamefont {Tan}},\ }\bibfield  {title}
  {\bibinfo {title} {{Pion Distribution Amplitudes from Functional QCD}},\
  }\href@noop {} {\bibfield  {journal} {\bibinfo  {journal} {in preparation}\ }
  (\bibinfo {year} {2026})}\BibitemShut {NoStop}%
\bibitem [{\citenamefont {Chang}\ \emph {et~al.}(2013)\citenamefont {Chang},
  \citenamefont {Cloet}, \citenamefont {Cobos-Martinez}, \citenamefont
  {Roberts}, \citenamefont {Schmidt},\ and\ \citenamefont
  {Tandy}}]{Chang:2013pq}%
  \BibitemOpen
  \bibfield  {author} {\bibinfo {author} {\bibfnamefont {L.}~\bibnamefont
  {Chang}}, \bibinfo {author} {\bibfnamefont {I.~C.}\ \bibnamefont {Cloet}},
  \bibinfo {author} {\bibfnamefont {J.~J.}\ \bibnamefont {Cobos-Martinez}},
  \bibinfo {author} {\bibfnamefont {C.~D.}\ \bibnamefont {Roberts}}, \bibinfo
  {author} {\bibfnamefont {S.~M.}\ \bibnamefont {Schmidt}},\ and\ \bibinfo
  {author} {\bibfnamefont {P.~C.}\ \bibnamefont {Tandy}},\ }\bibfield  {title}
  {\bibinfo {title} {{Imaging dynamical chiral symmetry breaking: pion wave
  function on the light front}},\ }\href
  {https://doi.org/10.1103/PhysRevLett.110.132001} {\bibfield  {journal}
  {\bibinfo  {journal} {Phys. Rev. Lett.}\ }\textbf {\bibinfo {volume} {110}},\
  \bibinfo {pages} {132001} (\bibinfo {year} {2013})},\ \Eprint
  {https://arxiv.org/abs/1301.0324} {arXiv:1301.0324 [nucl-th]} \BibitemShut
  {NoStop}%
\end{thebibliography}%

\end{document}